\documentclass[%
 reprint,
superscriptaddress,
nofootinbib,
 amsmath,amssymb,
aps,
prc,
floatfix,
]{revtex4-2}

\usepackage{graphicx}% Include figure files
\usepackage{dcolumn}% Align table columns on decimal point
\usepackage{bm}% bold math
\usepackage{hyperref}% add hypertext capabilities
\graphicspath{{./Figures/}}
\newcommand\blfootnote[1]{%
  \begingroup
  \renewcommand\thefootnote{}\footnote{#1}%
  \addtocounter{footnote}{-1}%
  \endgroup
}

\begin{document}

\preprint{APS/123-QED}

\title{IMSRG with flowing 3 body operators, and approximations thereof}% Force line breaks with \\

\author{S.~R.~Stroberg}
\email{rstroberg@nd.edu}
\affiliation{Department of Physics and Astronomy,
   University of Notre Dame,
   Notre Dame, IN, 46556, USA
}
 \affiliation{Physics Division,
 Argonne National Laboratory,
 Lemont, IL, 60439, USA}
\affiliation{
   Department of Physics,
   University of Washington,
   Seattle, WA 98195, USA
 }
\author{T.~D.~Morris}%
\affiliation{Quantum Information Science Section, 
Oak Ridge National Laboratory, 
Oak Ridge, TN 37831, USA}

\author {B.~C.~He}%
\affiliation{Department of Physics and Astronomy,
   University of Notre Dame,
   Notre Dame, IN, 46556, USA
}
\blfootnote{This manuscript has been authored by UT-Battelle, LLC under Contract No. DE-AC05-00OR22725 with the U.S. Department of Energy. The United States Government retains and the publisher, by accepting the article for publication, acknowledges that the United States Government retains a non-exclusive, paid-up, irrevocable, world-wide license to publish or reproduce the published form of this manuscript, or allow others to do so, for United States Government purposes. The Department of Energy will provide public access to these results of federally sponsored research in accordance with the DOE Public Access Plan (http://energy.gov/downloads/doe-public-access-plan).}
\date{\today}% It is always \today, today,
             %  but any date may be explicitly specified

\begin{abstract}
We explore the impact of retaining three-body operators within the in-medium similarity renormalization group (IMSRG), as well as various approximations schemes.
After studying two toy problems, idential fermions with a contact interaction and the Lipkin-Meshkov-Glick model, we employ the valence-space formulation of the IMSRG to investigate the even-$A$ carbon isotopes with a chiral two-body potential.
We find that retaining only those commutators expressions that scale as $N^7$ provides an excellent approximation of the full three-body treatment.
\end{abstract}

%\keywords{Suggested keywords}%Use showkeys class option if keyword
                              %display desired
\maketitle

%\tableofcontents

\section{\label{sec:Intro}Introduction}

There has been considerable progress in the past few years in ab initio many-body methods for atomic nuclei~\cite{Hergert2020}, with an explosion in the number of available methods, as well as in the systems amenable to ab initio treatment.
At present, nuclei from $A=2$ up to $A\sim 208$ can be accessed~\cite{Miyagi2022,Hu2022,Hebeler2023}, including many open-shell nuclei~\cite{Hergert2013,Soma2014,Sun2018,Stroberg2021,Tichai2023}.
There has simultaneously been great progress in quantifying uncertainties due to the truncation of effective field theory (EFT) to a finite order, especially the use of emulators for the solution of the many-body problem~\cite{Frame2018,Ekstrom2019,Melendez2019,Jiang2022,Belley2024}.
One of the most important remaining open issues in ab initio nuclear theory is a robust and reliable assessment of the error due to approximations made in solving the many-body problem.

In this paper, we focus on the in-medium similarity renormalization group (IMSRG) approach~\cite{Tsukiyama2011,Tsukiyama2012,Hergert2016,Hergert2016PS,Stroberg2019}, and work towards understanding how the adopted truncation scheme manifests as errors in the predicted observables.
The paper is organized as follows.
In section \ref{sec:IMSRG}, we summarize the IMSRG method and discuss the truncation scheme.
In section~\ref{sec:approximations} we discuss approaches to approximate a full treatment of three-body operators within the IMSRG.
In section~\ref{sec:benchmarking} we explore two toy models to develop intuition of the IMSRG truncation scheme, before applying the machinery to a more realistic calculation of the carbon isotopes.
Finally, in section~\ref{sec:cost} we briefly discuss the computational cost of our implementation of the IMSRG.

\section{In-medium similarity renormalization group\label{sec:IMSRG}}
In this section, we briefly recapitulate the formalism of the in-medium similarity renormalization group, which has previously been summarized in several review papers, including~\cite{Hergert2016,Stroberg2019}.
%(Readers familiar with the formalism, as presented in Refs. \cite{Hergert2016,Stroberg2019} can safely skip to section~\ref{sec:Truncation}).
\subsection{Basic ingredients\label{sec:basic}}
The basic approach of the SRG is to perform a unitary transformation on the Hamiltonian 
\begin{equation}\label{eq:Hsdef}
    H(s) \equiv U(s) H U^{\dagger}(s),
\end{equation}
in order to bring it to a form which is more amenable to solving the many-body Schr\"odinger equation.
The dependence of the unitary transformation $U(s)$ on the flow parameter $s$ is specified by the generator $\eta$
\begin{equation}
    \frac{d}{ds}U(s) = \eta(s)U(s).
\end{equation}
To ensure that $U(s)$ is unitary, we require that $\eta$ is anti-hermitian, $\eta^{\dagger}=-\eta$.
This leads to the SRG flow equation
\begin{equation}\label{eq:dHds}
    \frac{d}{ds}H(s) = [\eta(s),H(s)].
\end{equation}
Integration of the ordinary differential equation \eqref{eq:dHds} yields $H(s)$.
Given that the only requirement on $\eta$ is that it is anti-hermitian, there is some art in choosing a generator which efficiently leads to an improved form of the Hamiltonian.
Here, we use the White generator~\cite{White2002}, or its arctangent variant which we write schematically as
\begin{equation}\label{eq:generators}
    \eta^{\rm Wh} = \frac{H^{\rm od}}{\Delta}
    \hspace{1em},\hspace{2em}
    \eta^{\rm atan} = \frac{1}{2} {\rm atan}\left(\frac{2H^{\rm od}}{\Delta} \right).
\end{equation}
In \eqref{eq:generators} $H^{\rm od}$ is the suitably-defined off-diagonal part of the Hamiltonian (the part which should be suppressed by the RG evolution).
$\Delta$ is an energy denominator, typically given by Epstein-Nesbet or M{\o}ller-Plesset partitioning.
We employ a superoperator notation; the division and arctangent should be understood as acting elementwise.
For a more explicit description, see e.g.~\cite{Hergert2017a}.

\subsection{Magnus formulation\label{sec:Magnus}}
Rather than directly integrating \eqref{eq:dHds}, it is often convenient both numerically and formally to use the Magnus formulation of the SRG~\cite{Morris2015}.
In this case, we express the unitary transformation as an exponentiated anti-hermitian operator $\Omega$
\begin{equation}\label{eq:Umagnus}
U(s)=e^{\Omega(s)}.
\end{equation}
We may derive a flow equation for $\Omega(s)$ by using the Baker-Campbell-Hausdorff formula
\begin{equation}\label{eq:dOmegads}
    \frac{d}{ds}\Omega(s) = \sum_{k=0}^{\infty} \frac{B_k}{k!} [\Omega(s),\eta(s)]^{(k)}
\end{equation}
where $B_k$ is the $k$th Bernoulli number and $[\eta,\Omega]^{(k)}$ indicates a $k$-fold nested commutator
\begin{equation}\label{eq:defnestedcomm}
[\Omega,\eta]^{(k)} =
\begin{cases}
    [\Omega,[\Omega,\eta]^{(k-1)}] & k>0 \\
   \eta & k=0
 \end{cases}.
\end{equation}
The flowing Hamiltonian is then given by
\begin{equation}\label{eq:magnusH}
    H(s) = e^{\Omega(s)}H(0)e^{-\Omega(s)} = \sum_{k=0}^{\infty} \frac{1}{k!} [\Omega(s),H(0)]^{(k)}.
\end{equation}

In order to perform a calculation, we must choose a representation for our operators.
Typically we choose a Fock space representation, writing operators in terms of strings of creation and annihilation operators.
For example, a particle-number conserving operator $O$ would be written as
\begin{equation}
\label{eq:fockspaceA}
\begin{aligned}
    O = O_0 &+ \sum_{ij} O_{ij} \{a^{\dagger}_ia_j\}
   + \frac{1}{(2!)^2} \sum_{ijkl}O_{ijkl}\{a^{\dagger}_ia^{\dagger}_ja_la_k\} \\
   &+ \frac{1}{(3!)^2}\sum_{ijklmn} O_{ijklmn} \{a^{\dagger}_ia^{\dagger}_ja^{\dagger}_ia_ma_na_l\}+\ldots
   \end{aligned}
\end{equation}
where the ellipses indicate terms involving more than three creation and three annihilation operators.
The braces $\{\cdot \}$ indicate that the the string of creation and annihilation operators are normal ordered with respect to some reference state $|\Phi\rangle$ (which could possibly be the true vacuum $|0\rangle$) so that $\langle \Phi| \{a^{\dagger}\ldots a \}|\Phi\rangle=0$.
The coefficients $O_{ijkl},O_{ijklmn}$ are antisymmetrized matrix elements so that, e.g., $O_{ijkl}=\langle ij | O |kl \rangle $, where $|kl\rangle=a^{\dagger}_ka^{\dagger}_l|0\rangle$.

Occasionally, we find that the norm $\|\Omega\|$ becomes large during the IMSRG flow and evaluating \eqref{eq:dOmegads} and \eqref{eq:magnusH} requires the evaluation of many nested commutators before convergence is reached.
In this case it becomes computationally advantageous to split up the transformation so that \eqref{eq:Umagnus} becomes
\begin{equation}
    U(s) = U(s-s_1)U(s_1) = e^{\Omega(s-s_1)}e^{\Omega(s_1)}.
\end{equation}
We then save the intermediate Hamiltonian $H(s_1)$ and evaluating $H(s)=e^{\Omega(s-s1)}H(s_1)e^{-\Omega(s-s1)}$
converges more rapidly.
If $\Omega(s-s_1)$ becomes large, we can repeat the procedure so that after $n$ splits,
\begin{equation}
    U(s) = e^{\Omega(s-s_n)}\ldots e^{\Omega(s_2-s_1)} e^{\Omega(s_1)}.
\end{equation}
If we are only interested in energies, we only need to retain the most recent $H$ and $\Omega$ operators.
In the limit that we split after each infinitessimal step $ds$, the procedure becomes equivalent to directly integrating the flow equation \eqref{eq:dHds}.

For certain approximation schemes, such as the IMSRG(2*) approximation described below, this splitting leads to an enhanced error due to missing cross terms.
It is then useful to adopt a strategy which we call ``hunter-gatherer'', in which we split the transformation into two parts 
\begin{equation}
    U(s)=e^{\Omega_G(s)}e^{\Omega_{H}(s)}
\end{equation}
where $\Omega_{H}$ is the ``hunter'' and $\Omega_{G}$ is the ``gatherer''.
Whenever the norm $\| \Omega_{H}\|$ gets beyond a (typically small) threshold, we update the gatherer via the Baker-Campbell-Hausdorff formula
\begin{equation}
    \Omega_G \to \sum_{k=0}^{\infty} \frac{B_k}{k!}[\Omega_{G},\Omega_H]^{(k)}.
\end{equation}
This approach retains the benefits of the aforementioned splitting strategy, without losing the cross terms in the IMSRG(2*) approximaiton.

\subsection{Truncation of the SRG flow equations and the \texorpdfstring{NO$n$B}{NOnB} approximation\label{sec:Truncation}}

From \eqref{eq:fockspaceA} combined with either \eqref{eq:dHds} or \eqref{eq:dOmegads} and \eqref{eq:magnusH}, we can see that the main computational task required for an IMSRG calculation is the evaluation of commutators of Fock-space operators.
Even if $\eta$ and $H$ are initially two-body operators, integrating \eqref{eq:dHds}, or evaluating the nested commutators in \eqref{eq:defnestedcomm} and \eqref{eq:magnusH} will induce three-body and higher-body operators.
To make the calculation feasible, all operators, including intermediate commutator expressions, are truncated at the two-body level, yielding the IMSRG(2) approximation\footnote{In free-space SRG calculations, due to the translation-invariance of the reference $|\Phi\rangle = |0\rangle$, three-body operators can be managed more efficiently by eliminating the center of mass coordinate and working in Jacobi coordinates~\cite{Jurgenson2009,Roth2014}).}.

The IMSRG(2) approximation can be viewed as repeatedly making the normal-ordered two-body (NO2B) approximation, which is also used with other methods, and which is very accurate beyond $A\sim 16$~\cite{Hagen2014,Roth2012,Djarv2021}.
%\srs{Worth citing \cite{Drell1953} for arguments about convergence in many-body forces?}
For a single-determinant reference $|\Phi\rangle$, the normal-ordered $n$-body (NO$n$B) approximation becomes exact when $n$ is equal to the maximum number of quasi-particles (particles plus holes) present in the exact wave function.
For example, a one-particle-one-hole (two quasi-particle) configuration receives contributions from a normal-ordered two-body operator;
but a normal ordered three-body operator will not contribute:
$\langle \Phi^a_i | \{a^{\dagger}a^{\dagger}a^{\dagger}aaa \} | \Phi^b_j \rangle =0$.
Likewise, a two-particle-two-hole configuration receives contributions from a four-body operator, but not a five-body operator.

For an $A$-body system with a single-determinant reference with $A$ particles, at most $A$-particle-$A$-hole configurations can be present in the wave function, and so the NO$n$B approximation is exact for $n\geq 2A$~\cite{Kehrein2006}.
With a vacuum reference, at most $A$ quasiparticles are present, and so the NO$n$B approximation is exact for $n\geq A$.
We are interested in treating systems with $A\gg 3$, and going beyond the NO3B approximation is generally prohibitively expensive, so the condition $n\ll A$ is inevitable.

The NO$n$B approximation will also be exact if there are no $m$-body operators with $m>n$;
for example, working at low order in chiral effective field theory~\cite{Machleidt2011,Epelbaum2009} there are no operators with $m>3$, so the NO3B approximation is exact.
However, SRG evolution will induce operators with $m>3$ so IMSRG(3) is no longer exact.
The error due to neglecting $m$-body operators will depend on the relative importance of $m$-quasiparticle configurations in the wave function.
For the SRG with a vacuum reference, all configurations have $A$ quasiparticles, while for IMSRG (with a judicious choice of reference) the wave function should be dominated with $m$-quasiparticle configurations with $m\ll A$.

Focusing for concreteness on the NO2B approximation of a three-body operator, we schematically write the expectation value of a string of creation/annihilation operators in the exact wave function $|\Psi\rangle$, with  reference $|\Phi\rangle$, as
\begin{equation}\label{eq:wick3b}
\begin{aligned}
    \langle \Psi | a^{\dagger} a^{\dagger}a^{\dagger} aa a| \Psi \rangle \sim~& 
    \langle \Phi | a^{\dagger} a^{\dagger} a^{\dagger} aa a | \Phi \rangle \\
    +&  \langle \Phi | a^{\dagger} a^{\dagger} a a| \Phi \rangle  ~ 
    \langle \Psi |\{ a^{\dagger}   a \}| \Psi \rangle \\
    +&  \langle \Phi | a^{\dagger}   a| \Phi \rangle   ~
    \langle \Psi |\{ a^{\dagger} a^{\dagger} a a \}| \Psi \rangle\\
    +&
    \langle \Psi |\{ a^{\dagger} a^{\dagger}a^{\dagger} aa a \}| \Psi \rangle.
    \end{aligned}
\end{equation}
The first term in \eqref{eq:wick3b} will scale as $A^3$.
The second term will scale as $A^2 N_q$, where $N_q$ is the number of quasiparticles~\footnote{Here, we assume that the size of $\langle \Psi|\{a^{\dagger}a \}|\Psi\rangle$ is reasonably approximated by the diagonal terms $N_q = \sum_p \langle p^{\dagger}p\rangle + \sum_h \langle h h^{\dagger}\rangle$. In nuclear structure applications, many of the off-diagonal terms will be suppressed by selection rules on the quantum numbers $j,\pi,t_z$.}.
The third term will go as $A N_q^2$, and the fourth as $N_q^3$, so we see that we have an organization in powers of $N_q/A$~\cite{Kriel2005,Friman2011}.
The approximation is improved by the IMSRG flow, which suppresses excitations so that $N_q(s)\to 0$ as $s\to \infty$.
 This argument is complicated somewhat by the fact that the three-body operators induced by the IMSRG flow are not uniformly distributed over all orbits, but instead are concentrated near the Fermi surface.
 Nevertheless, so long as the typical number of quasiparticles is smaller than the number of particles feeling the induced forces, we expect the effect of many-body operators to be suppressed.

%\begin{figure*}
\begin{figure}
\centering
\includegraphics[width=1.0\columnwidth]{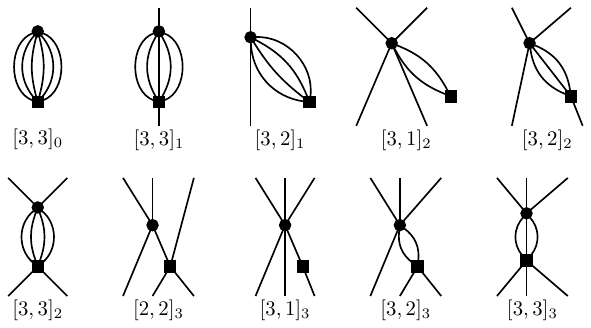}
%\setlength{\tabcolsep}{14pt}
%%\begin{tabular}{cccccccccc}
%\begin{tabular}{ccccc}
%\input{Diagrams/skeleton_330.tikz} &
%\input{Diagrams/skeleton_331.tikz} &
%\input{Diagrams/skeleton_321.tikz} &
%\input{Diagrams/skeleton_312.tikz} &
%\input{Diagrams/skeleton_322.tikz} \\
%$[3,3]\to 0$ &
%$[3,3]\to 1$ &
%$[3,2]\to 1$ &
%$[3,1]\to 2$ &
%$[3,2]\to 2$ \\[5pt]
%%
%\input{Diagrams/skeleton_332.tikz} &
%\input{Diagrams/skeleton_223.tikz} &
%\input{Diagrams/skeleton_313.tikz} &
%\input{Diagrams/skeleton_323.tikz} &
%\input{Diagrams/skeleton_333.tikz} \\
%$[3,3]\to 2$ &
%$[2,2]\to 3$ &
%$[3,1]\to 3$ &
%$[3,2]\to 3$ &
%$[3,3]\to 3$ \\
%\end{tabular}
    \label{fig:imsrg3skeletons}
    \caption{Hugenholtz skeleton diagrams indicating the commutator topologies present in IMSRG(3) which are omitted in the IMSRG(2) approximation. The circle and square represent the two operators entering into the commutator.}
\end{figure}
%\end{figure*}

\section{\label{sec:approximations}Approximations to full IMSRG(3)}

The full commutator expressions necessary for IMSRG(3) in the uncoupled and $J$-coupled representations are presented in Appendices~\ref{app:commutators_m} and \ref{app:commutators_j} respectively (see also refs~\cite{Tsukiyama2010,Morris2016,Hergert2016,Heinz2021}).
The corresponding Hugenholtz skeleton diagrams which enter at the IMSRG(3) level are presented in Fig.~\ref{fig:imsrg3skeletons}.
The naive computational scaling to evaluate a particular term can be read off of the diagram by counting the number of fermion lines.
For example, the first diagram in Fig.~\ref{fig:imsrg3skeletons}, labeled\footnote{Here we use the shorthand notation $[a,b]_c$ for the $c$-body piece of the commutator of an $a$-body operator with a $b$-body operator.} $[3,3]_0$, has six fermion lines and thus scales as $N^6$, where $N$ is the number of single-particle states in the basis.
The final diagram in Fig.~\ref{fig:imsrg3skeletons} involves nine fermion lines and thus scales as $N^9$.
This is too expensive for realistic applications (see section~\ref{sec:cost}), and so we shall explore some approximation schemes which can render the calculation more tractable.

\subsection{Goose-tank diagrams and the IMSRG(\texorpdfstring{2$^*$}{2*}) approximation}
\begin{figure}
    \centering
    \includegraphics[width=0.5\columnwidth]{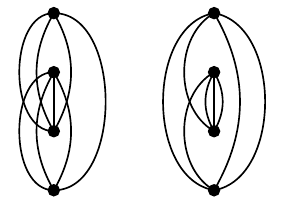}
    \caption{Hugenholtz skeleton diagrams indicating the fourth-order quadruples which are undercounted in the IMSRG(2) (cf Fig 5.5 of~\cite{Shavitt2009}).
    }
    \label{fig:quadruples}
\end{figure}

In refs.~\cite{Morris2016,Hergert2016}, a perturbative analysis of the IMSRG demonstrated that the IMSRG(2) ground state energy is exact through third order in MBPT, while some fourth-order and higher-order terms are missed.
Specifically, all fourth-order diagrams with three-particle-three-hole intermediate states (triples) are missed, while a subset of diagrams with four-particle-four-hole intermediate states (asymmetric quadruples, shown in Fig.~\ref{fig:quadruples}) are undercounted by a factor $1/2$.
Coupled cluster with singles and doubles (CCSD) also misses the fourth-order triples, but correctly includes the quadruples~\cite{Morris2016}.

\begin{figure}[t]
   \includegraphics[width=0.4\columnwidth]{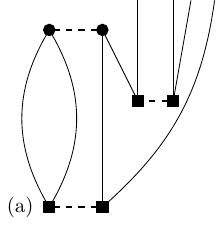}
   \hspace{3mm}
   \includegraphics[width=0.4\columnwidth]{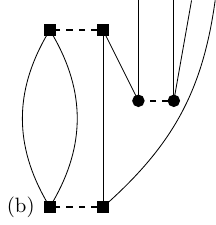}\\
\vspace{2mm}
   \includegraphics[width=0.35\columnwidth]{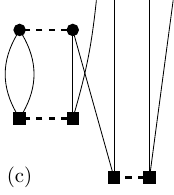}
   \hspace{10mm}
   \includegraphics[width=0.35\columnwidth]{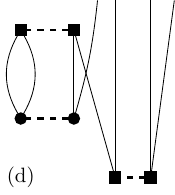}
    \caption{(a) and (b): Diagrammatic illustration of a contributions to $[\eta,[\eta,H]_{3}]_2$, which we denote ``goose-tank'' diagrams. (c) Diagram arising from $[\eta,[\eta,H]_1]_2$ which gives a contribution identical to that of (a).
    %(d) Diagram equivalent to (b) which does not arise in the IMSRG.
    (d) Diagram arising from $[\eta,[\eta,H]_1]_2$ which is approximately equal to (b).  In these diagrams, vertices of $H$ are represented by circles, while vertices of $\eta$ are represented by squares.}
    \label{fig:goosetank}
\end{figure}

Within the IMSRG, the energy diagrams in Fig.~\ref{fig:quadruples} are obtained\footnote{The commutator of the third-order generator with the first order vertex yields an identical expression\cite{Morris2016,Hergert2016}.} from the commutator of the third order two-body vertex $\Gamma^{[3]}$ with the first-order generator $\eta^{[1]}$.
The missing half of the quadruples diagrams are due to the contributions to $\Gamma^{[3]}$ which involve an intermediate three-body operator.
These are proportional to $[\eta,[\eta,\Gamma]_{\rm 3b}]_{\rm 2b}$, as indicated in Figs.~\ref{fig:goosetank}(a) and \ref{fig:goosetank}(b), where $\eta$ and $\Gamma$ are evaluated at $s=0$.
We refer to these diagrams as ``goose-tank'' diagram because of their appearance when rotated 90 degrees.
We note that their importance was also identified by Evangelista and collaborators in the context of quantum chemistry~\cite{Evangelista2012,Evangelista2014,Li2020}.

The half of the asymmetric quadruples which are included in the IMSRG(2) arise from contributions to $\Gamma^{[3]}$ proportional to $[\eta,[\eta,\Gamma]_{\rm 1b}]_{\rm 2b}$, involving an intermediate one-body operator, as indicated in Figs.~\ref{fig:goosetank}(c) and \ref{fig:goosetank}(d).
Evaluation of the diagrams in Fig.~\ref{fig:goosetank} reveals that \ref{fig:goosetank}(c) is equal to~\ref{fig:goosetank}(a), while~\ref{fig:goosetank}(d) is only approximately equal to~\ref{fig:goosetank}(b).
Nevertheless, \ref{fig:goosetank}(d) and~\ref{fig:goosetank}(b) give identical contributions to the fourth-order energy when further contracted with the generator $\eta$ to yield a zero-body operator.
The full fourth-order quadruples can therefore be restored by modifying the IMSRG(2) to the so-called IMSRG(2$^*$) scheme.
In the standard formulation, we modify the flow equation to be
\begin{equation} \label{eq:flow2star}
    \frac{d}{ds}H(s) = [\eta(s),H(s)+\chi(s)]
\end{equation}
where $\chi(s)$ is an auxiliary one-body operator
%, shown diagrammatically in Fig.~\ref{fig:chi},
which obeys the flow equation
\begin{equation}\label{eq:dchids}
    \frac{d}{ds}\chi_{ij} = (n_in_j + \bar{n}_i\bar{n}_j)[\eta_{\rm 2b},H_{\rm 2b}]_{ij}.
\end{equation}
In the Magnus formulation, the nested commutators used in \eqref{eq:magnusH} are modified to
\begin{equation}
    [\Omega,H]^{(k+1)} = \begin{cases}
        [\Omega,[\Omega,H]^{(k)}] & k=0 \\
        [\Omega,[\Omega,H]^{(k)}+\chi^{(k)}] & k>0,
    \end{cases} 
\end{equation}
and the one-body operator $\chi^{(k)}$ is defined as
\begin{equation}
    \chi^{(k)}_{ij} = (n_in_j + \bar{n}_i\bar{n}_j)
%    [\Omega_{\rm 2b},X^{k-1}_{\rm 2b}]_{ij}
    [\Omega_{\rm 2b},[\Omega,H]^{(k-1)}_{\rm 2b}]_{ij}.
\end{equation}
As noted in section~\ref{sec:Magnus}, the procedure is made more complicated if we split the transformation into multiple steps, and so it is best to use the hunter-gatherer scheme.

%\begin{figure}
%    \centering
%    \input{Diagrams/chi.tikz}
%    \caption{Topologies contributing to the auxiliary one-body operator $\chi$ used in the IMSRG(2*) scheme.}
%    \label{fig:chi}
%\end{figure}

In effect, the IMSRG(2$^*$) scheme approximates the goose-tank diagram in Fig.~\ref{fig:goosetank}(b) with a second copy of the diagram in Fig.~\ref{fig:goosetank}(d).
This replacement is exact for the fourth-order energy contribution, but only approximate for higher-order terms.
Explicitly, we have
\begin{equation}
    \begin{aligned}
%            \Gamma^{(a)}_{ijkl} &\sim \tfrac{1}{2}\sum_{abcd} \bar{n}_a\bar{n}_b n_c n_d\frac{\Gamma_{ijkd}\Gamma_{cdab}\Gamma_{abcl}}{\Delta_{abcl}\Delta_{ijdk}} \\
%        \Gamma^{(b)}_{ijkl} &\sim -\tfrac{1}{2}\sum_{abc}\bar{n}_a\bar{n}_b n_c n_d \frac{\Gamma_{ijkd}\Gamma_{cdab}\Gamma_{abcl}}{\Delta_{abcl}\Delta_{cdab}} \\
        \Gamma^{(b)}_{ijkl} &\sim \tfrac{1}{2}\sum_{abcd} \bar{n}_a\bar{n}_b n_c n_d\frac{\Gamma_{ijdk}\Gamma_{cdab}\Gamma_{abcl}}{\Delta_{cdab}\Delta_{abcl}} \\
        \Gamma^{(d)}_{ijkl} &\sim \tfrac{1}{2}\sum_{abc}\bar{n}_a\bar{n}_b n_c n_d \frac{\Gamma_{ijdk}\Gamma_{cdab}\Gamma_{abcl}}{\Delta_{cdab}\Delta_{ijdk}} \\
    \end{aligned}
\end{equation}
where $i,j$ are particle orbits and $k,l$ are holes.
Diagram (d) comes with a minus sign in~\eqref{eq:dchids} due to the commutator, so
IMSRG(2$^*$) approximately captures diagram (b) assuming $\Delta_{ijdk}\approx - \Delta_{cdab}$, which is true to the extent that all the two-particle-two-hole denominators $\Delta_{pp'hh'}$ are equal.
Because these goose-tank topologies are of pphh form, they feed back into the generator $\eta$ and have a non-linear effect beyond the fourth-order energy.
As we will see below, in some difficult cases in which the IMSRG(2) flow diverges, the IMSRG(2*) modification stabilizes the flow.
An alternative strategy, in which the relevant double commutators are evaluated exactly is presented in~\cite{He2024}.

\subsection{Perturbative triples\label{sec:perturbative-triples}}
The remaining fourth order energy diagrams, the triples, are produced in the IMSRG by terms involving the three-body generator~\cite{Morris2016,Hergert2016}.
After solving the flow equation~\eqref{eq:dOmegads} at the IMSRG(2) level, the transformed Hamiltonian is
\begin{equation}
    \bar{H} = e^{\Omega}He^{-\Omega} = \bar{H}_{\rm IMSRG(2)} + \bar{W}
\end{equation}
where $\bar{W}$ is the induced three-body interaction.
We can directly evaluate the contribution of this term to the ground state energy in second-order perturbation theory.
However, it is more convenient to use the Magnus formulation here, because it makes generalizations beyond perturbative counting and to the VS-IMSRG more straightforward.
We wish to suppress the off-diagonal part of $\bar{W}$, and the leading-order estimate of this, using the White generator, is
\begin{equation}
    \bar{\Omega}=\frac{\bar{W}^{\rm od}}{\bar{\Delta}}.
\end{equation}
We perform a second unitary transformation on the Hamiltonian yielding $\bar{\bar{H}}=e^{\bar{\Omega}}\bar{H} e^{-\bar{\Omega}}$.
The zero-body piece of this is
\begin{equation}
    \bar{\bar{H}}_{\rm 0b} = \bar{H}_{\rm 0b} + [\bar{\Omega},\bar{H}]_{\rm 0b} + \tfrac{1}{2}[\bar{\Omega},[\bar{\Omega},\bar{H}]]_{\rm 0b} +\ldots
\end{equation}
and the elipses indicate terms of order $\bar{\Omega}^3$ which we neglect here.
Assuming the denominator $\bar{\Delta}$ is chosen according the partitioning we use for perturbative counting (i.e. Moller-Plesset or Epstein-Nesbet), then taking just the zero-order piece of $\bar{H}$ gives
$[\bar{\Omega},\bar{H}_{\rm 1b,2b}^{[0]}]=-\bar{W}$
so
%
%Since the one and two-body parts of $\bar{H}$ are already diagonal, and assuming $\bar{\Delta}$ is formed from the diagonal part of $\bar{H}$, we have $[\bar{\Omega},\bar{H}_{\rm 1b,2b}]=-\bar{W}$, so
\begin{equation}\label{eq:Hbarbar}
    \bar{\bar{H}}_{\rm 0b} = \bar{H}_{\rm 0b} + \tfrac{1}{2}[\bar{\Omega},\bar{W}]_{\rm 0b} +\ldots
\end{equation}
that is, we should take half the value of the $[3,3]_0$ commutator.
%\footnote{Here we introduce the shorthand notation $[a,b]_c$ for the $c$-body piece of the commutator of an $a$-body operator with a $b$-body operator.}.
Taking the explicit form of the commutator, we have the perturbative correction to the energy~\cite{Morris2016}
%Terms involving $[\eta_{\rm 3b},[\eta_{\rm 2b},\Gamma]_{\rm 3b}]$ and those involving $[\eta_{\rm 2b},[\eta_{\rm 3b},\Gamma_{\rm 2b}]]$ give identical contributions to the fourth-order energy.
% Unlike the IMSRG(2$^*$) approximation, the evaluation of these commutators scales as $N^7$.
% We can ensure that the energy is exact through fourth order by including this contribution perturbatively as~\cite{Morris2016}
 \begin{equation}\label{eq:pertriples}
     \Delta E_{[3]} = \frac{1}{(3!)^2} \sum_{abcijk} 
     \bar{n}_a\bar{n}_b\bar{n}_cn_i n_j n_k
    \frac{\bar{W}_{ijkabc}\bar{W}_{abcijk}}{\bar{\Delta}_{ijkabc}}.
 \end{equation}
 There are several possible ways to approximate $\bar{W}$ and $\bar{\Delta}$~\cite{Morris2016} which only involve a single $N^7$ commutator evaluation\footnote{More precisely, this is $N_h^3N_p^4$, which is better than the worst IMSRG(2) commutators.}, and which differ in the content beyond fourth order.
 A strictly fourth-order correction would be $\overline{W}=[\eta(0),\Gamma(0)]_{\rm 3b}$, with $\bar{\Delta}$ taken from $H(0)$.
 In the Magnus formulation, we can just as easily take the full $\Omega(\infty)$ obtained from the IMSRG(2) or IMSRG(2$^*$) solution.
 In fact, we can also include higher nested commutators by using an intermediate
 \begin{equation}\label{eq:Wbar}
     \bar{W} = \sum_{k=1}^{\infty}\frac{1}{k!}[\Omega,H]^{(k)}_{\rm 3b}
     = [\Omega,\widetilde{H}]_{\rm 3b}
 \end{equation}
 with
 \begin{equation}
     \widetilde{H} \equiv \sum_{k=0}^{\infty}\frac{1}{(k+1)!}[\Omega,H]^{(k)}.
 \end{equation}
 In this paper we will take the definition \eqref{eq:Wbar} and M{\o}ller-Plesset energy denominators obtained from the one-body part of $H(\infty)$.
 We will indicate this approximation Magnus(2$^{*}$)[3].
 
In the valence-space formulation, the connection to perturbation theory is formally less straightforward, but~\eqref{eq:Hbarbar} remains valid.
We can still use~\eqref{eq:pertriples} to approximately capture the leading effects of the induced three-body interaction.
This effectively performs second-order perturbation theory on the (ensemble) reference state.
A more formally correct choice would be to use the off-diagonal induced three-body interaction which connects valence configurations to non-valence configurations, namely $W^{\rm od}\in \{W_{pppccc},W_{ppqccv},W_{ppqcvv},W_{ppqvvv}\}$, where $c,v,q$ indicate core, valence, and excluded orbits, respectively, and $p\in\{q,v\}$.
The part of $\bar{\bar{H}}$ acting purely in the valence space, including the induced 3N interaction, should then be included in the valence space diagonalization (possibly perturbatively).
The restriction of external lines to valence orbits would keep the calculation tractable.
This direction will be left for a future study.

\subsection{Truncation to \texorpdfstring{$N^7$}{N7} scaling terms}
Alternatively, we may assume that not all topologies contributing to the full IMSRG(3) commutator expression are equally important.
A pragmatic way to approximate the full expression is to retain only those terms which scale as $N^7$ or better.
Explicitly, this amounts to neglecting the following topologies shown in Fig.~\ref{fig:imsrg3skeletons}:
$[3,3]_2$, $[3,2]_3$, and $[3,3]_3$.
In this approximation, errors made in the zero-body piece of $H(s)$ will be at least fifth order in the potential, errors to the one-body and two-body parts will be at least fourth order, and the error in the three-body part will be at least third order.
This truncation scheme was previously explored in Refs.~\cite{StrobergTalk2020,Heinz2021}, and we will refer to it as IMSRG(3N7).
In principle we can also consider terms scaling as $N^8$ or better, but in practice we find these extra terms are not worth the substantial computational effort required for direct evaluation.

\section{\label{sec:benchmarking}Benchmarking}

Before considering a more realistic case, we explore exactly-solvable simplified models which separately probe the short-range and long-range correlations present in a nucleus.

%%%%%%%%%%%%%%%%%%%%%%%%
\subsection{Two fermions with a contact interaction}
%%%%%%%%%%%%%%%%%%%%%%%%
The simplest non-trivial many-body system consists of two particles.
As discussed in Section~\ref{sec:Truncation}, IMSRG(2) and even IMSRG(3) are not exact for this system, because four-quasiparticle excitations are possible.
Therefore, we can gain some insight by considering a toy problem of two neutrons in a harmonic trap with a contact interaction.
We work in a basis spanned by three major harmonic oscillator shells (i.e. $0s,0p,1s0d$).
%[Make statement about how things change as the space is increased.]
The Hamiltonian is
\begin{equation}
    H = \sum_i \epsilon_i a^{\dagger}_ia_i + \tfrac{1}{4}\sum_{ijkl} V_{ijkl} a^{\dagger}_ia^{\dagger}_ja_la_k
\end{equation}
where the single-particle energy is $\epsilon_i=2n_i+\ell$ (we measure energy in terms of the oscillator energy $\hbar\omega$) and the potential is
\begin{equation} \label{eq:Vcontact}
    V(\vec{r}) = \frac{g}{({2\pi})^{3/2}} \delta(\vec{r}).
%    V(\vec{r}) = g\frac{b^3}{16\sqrt{\pi}} \delta(\vec{r})
\end{equation}
The relative coordinate $\vec{r}$ is expressed in units of the oscillator length $b=\sqrt{\hbar/m\omega}$.
With this definition, $g<0$ corresponds to an attractive interaction, and the normalization is chosen so that $\langle 0s 0s|V|0s 0s\rangle = g$.
%Matrix elements in the $J=0$ channel are listed in Table~\ref{tab:Vcontact} in Appendix~\ref{app:contact}.

%\begin{table}[]
%    \centering
%    \caption{Normalized, antisymmetrized two-body matrix elements $\langle aa |V | bb\rangle$ of the contact interaction~\eqref{eq:Vcontact} with $g=1$, coupled to $J=0$. By hermiticity the lower triangle is equal to the upper triangle and is not repeated.}
%    \label{tab:Vcontact}
%    \begin{ruledtabular}
%    \begin{tabular}{r  r | r r | r r r}
%              & $0s_{1}$ & $0p_{3}$ & $0p_1$ & $0d_{5}$ & $0d_{3}$ & $1s_{1}$  \\
%       $0s_1$ &  1.000  &  -0.707  &  -0.500  &   0.433 &   0.354 &   0.625  \\
%        \hline
%       $0p_3$ &   &   0.833  &   0.589  &  -0.714 &  -0.583 &  -0.324  \\
%       $0p_1$ &   &    &   0.417  &  -0.505 &  -0.413 &  -0.229  \\
%        \hline
%       $0d_5$ &   &    &    &   0.788 &   0.643 &   0.271  \\
%       $0d_3$ &   &    &    &    &   0.525 &   0.221  \\
%       $1s_1$ &   &    &    &    &    &   0.641  \\
%    \end{tabular}
%    \end{ruledtabular}
%\end{table}

While we must retain induced three-body (and four-body) terms in $H$ to obtain the exact result,
there are no three-particle or three-hole configurations possible for an $A=2$ system.
Only matrix elements involving $pph$ and $hhp$ configurations need to be considered.
Three-body terms in the generator, which are of the form $\eta_{ppphhh}$, will not contribute. 
Consequently, the perturbative triples discussed in section~\ref{sec:perturbative-triples} give no contribution to the energy, and the IMSRG(2*) approximation is exact through $g^4$ in perturbation theory.

\begin{figure}
    \centering
    \includegraphics[width=\columnwidth]{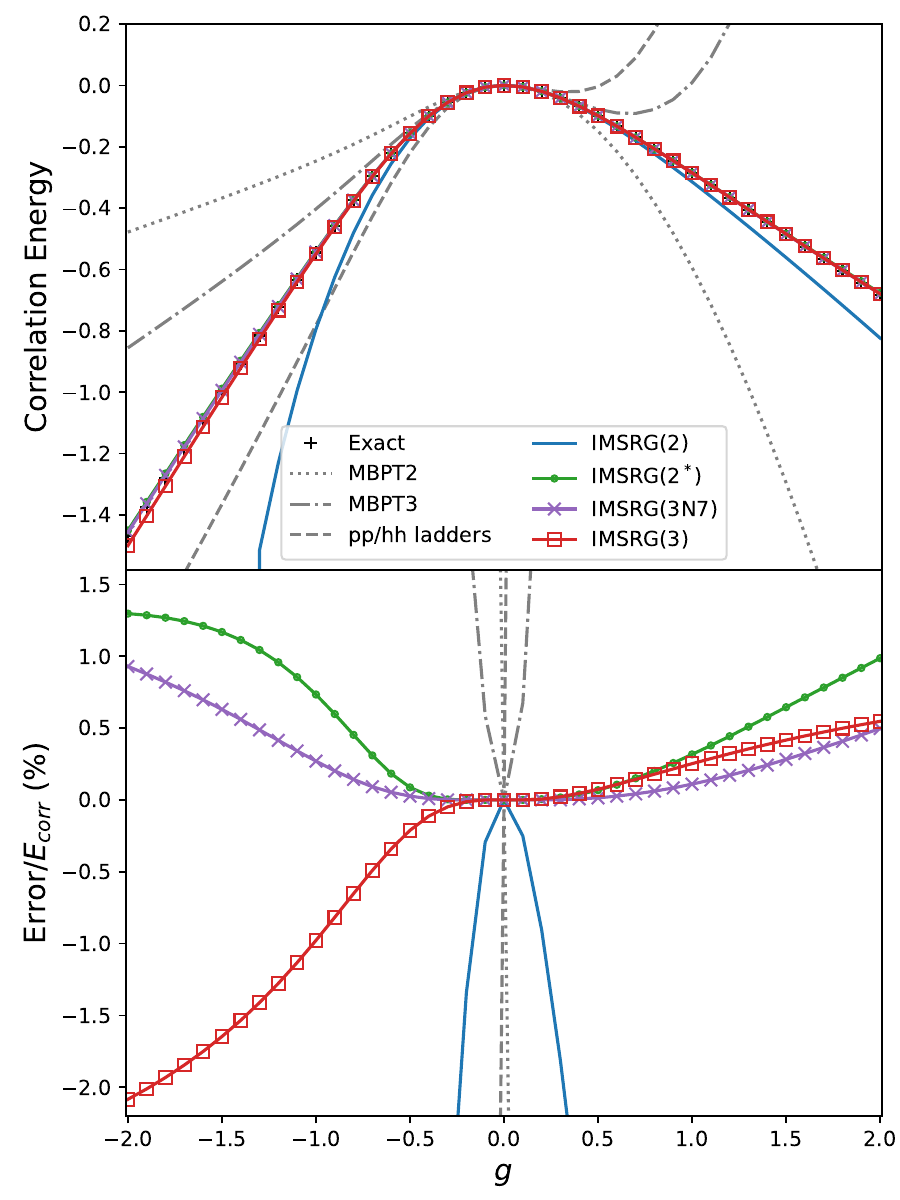}
    \caption{Correlation energy of two identical fermions in a harmonic trap with a contact interaction, obtained in various approximation schemes as a function of the coupling strength $g$. The lower panel shows the error in the energy as a percent of the correlation energy.}
    \label{fig:twofermions}
\end{figure}

Fig.~\ref{fig:twofermions} displays the ground state energy as a function of $g$, computed at various levels of approximation.
We include perturbation to second and third order, labeled MBPT(2) and MBPT(3), respectively, as well as MBPT(3) plus particle-particle and hole-hole ladders resummed to all orders.
We also include IMSRG(2), IMSRG(2*), IMSRG(3N7), and the full IMSRG(3).
In the top panel of Fig.~\ref{fig:twofermions}, we observe that for large positive coupling $g$, perturbation theory breaks down completely and resummed ladders do not fix the problem.
For attractive couplings, the relevant physics is captured by the ladder approximation.
In contrast, the IMSRG(2) is accurate for repulsive couplings, but diverges for sufficiently attractive coupling.
All approximations beyond IMSRG(2) are sufficiently accurate that they cannot be distinguished in the top panel.
In the bottom panel of Fig.~\ref{fig:twofermions}, we show the error relative to the exact solution, zoomed in to visualize the difference between the IMSRG approximations.
It appears that there is little further improvement in going from IMSRG(2*) to full IMSRG(3).

\begin{figure}
    \centering
    \includegraphics[width=\columnwidth]{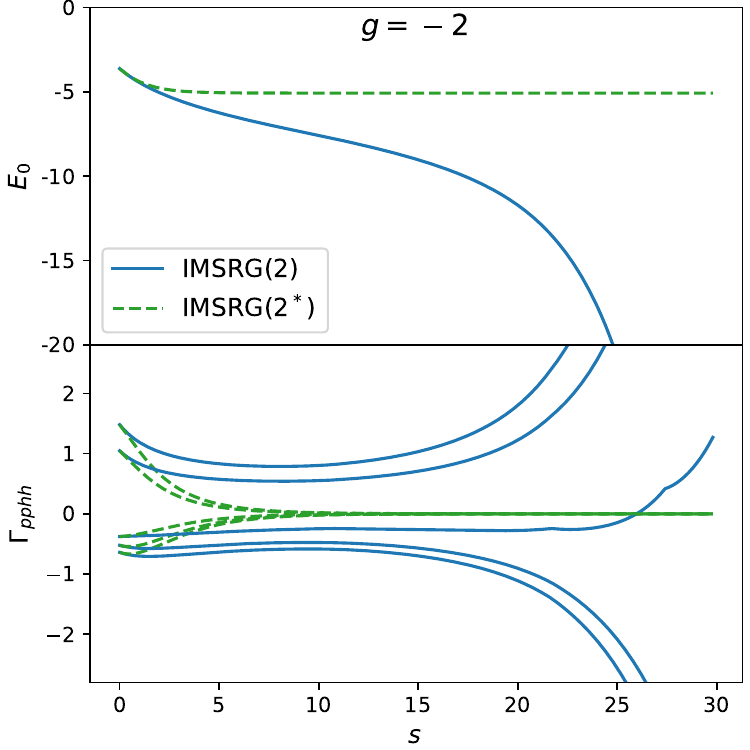}
    \caption{Flow of (a) zero-body term $E_0$ and (b) off-diagonal part of the two-body vertex $\Gamma_{pphh}$, for the two-fermion problem with $g=-2$, computed with the IMSRG(2) and IMSRG(2$^*$) approximations.}
    \label{fig:twofermionflow}
\end{figure}

\begin{figure}
    \centering
    \includegraphics[width=\columnwidth]{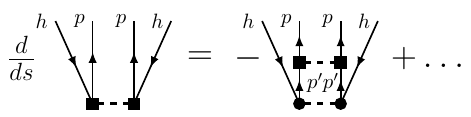}
    \caption{Diagram illustrating the commutator term $[\eta,\Gamma_{ppp'p'}]$ which contributes to the increasing magnitude of the off-diagonal two-body vertex $\Gamma_{pphh}$.}
    \label{fig:EtaV}
\end{figure}

 For $g\lesssim -1.8$, the IMSRG(2) flow diverges, while the flow converges to the fixed point for the IMSRG(2$^*$) and other approximations of IMSRG(3), as illustrated in Fig.~\ref{fig:twofermionflow}.
Fig.~\ref{fig:twofermionflow}(b), shows the flow of the off-diagonal part of the two-body vertex $\Gamma_{pphh}$.
With the IMSRG(2) approximation, the off-diagonal pieces are not suppressed, meaning $\frac{d}{ds}|\Gamma_{pphh}|>0$.
This can be traced back to the contribution illustrated in Fig.~\ref{fig:EtaV} in which the two-body vertex in the particle-particle channel $\Gamma_{ppp'p'}$ contracts with the generator $\eta_{p'p'hh}$ to give a contribution to $\frac{d}{ds}\Gamma_{pphh}$.
The contact interaction is approximately separable, i.e. $\Gamma_{abcd} \sim g F_{ab}F_{cd}$ for some one-body matrix $F$.
This leads to a coherent effect
\begin{equation}
\begin{aligned}
    \frac{d}{ds}\left( g F_{pp}F_{hh} \right) &\sim - \sum_{p'p'}\left( gF_{pp}F_{p'p'} \right)\left( gF_{p'p'} F_{hh} \right)\\
    &\sim (gF_{pp}F_{hh}) \cdot (-g) \cdot \sum_{p'p'}(F_{p'p'})^2.
    \end{aligned}
\end{equation}
For $g<0$ the contribution to $\frac{d}{ds}\Gamma_{pphh}$ has the same sign as $\Gamma_{pphh}(s)$, driving an exponential enhancement.

In contrast, the diagrams in Fig.~\ref{fig:goosetank} give a contribution with the opposite sign of $\Gamma_{pphh}$, leading to suppression.
This can be seen by considering the expression for the intermediate $\chi_{hh}$ as defined in~\eqref{eq:dchids}.
For the White generator this becomes
\begin{equation}
\begin{aligned}
    \chi_{hh} &= \sum_{p'} (\eta_{hhp'p'}\Gamma_{p'p'hh}-\Gamma_{hhp'p'}\eta_{p'p'hh}) \\
    &= -2\sum_{p'} \frac{|\Gamma_{p'p'hh}|^2}{\Delta_{p'p'hh}}
    \end{aligned}
\end{equation}
which is manifestly negative.
Contracting $\chi_{hh}$ with $\eta_{pphh}$ then yields a contribution $\frac{d}{ds}\Gamma_{pphh}\sim -\Gamma_{pphh}$, leading to suppression.

This illustrates how, in addition to restoring the fourth-order quadruples contribution to the ground state energy, the IMSRG(2$^*$) approximation---and the more expensive IMSRG(3N7) and IMSRG(3)---stabilizes the IMSRG flow in the non-perturbative regime.
For this toy problem, this constitutes the main effect of IMSRG(3).

\begin{figure}
    \centering
    \includegraphics[width=\columnwidth]{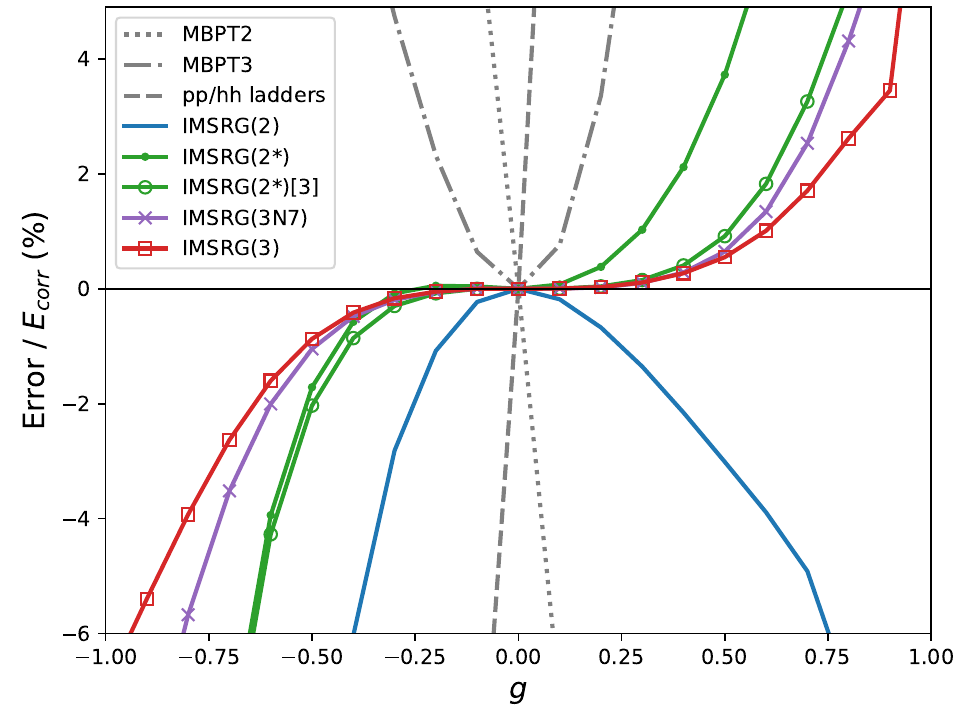}
    \caption{Error as a function of coupling constant $g$ for 8 identical fermions in a harmonic trap.}
    \label{fig:ContactN8}
\end{figure}

In order to illustrate the impact of the perturbative triples correction, we repeat the calculation with 8 particles in the trap.
The error as a function of the coupling $g$ is shown in Fig.~\ref{fig:ContactN8}.
The behavior is qualitatively similar to the two-particle case, with the exception that the perturbative triples now have an effect.
For repulsive $g>0$, the perturbative triples correction is substantial and accounts for nearly all of the difference between IMSRG(2$^*$) and IMSRG(3N7).
On the other hand, for $g<0$, IMSRG(2$^*$) overbinds relative to IMSRG(3N7); it is clear that the perturbative triples correction, which necessarily lowers the energy, cannot correct for this difference.
%We see that the IMSRG(2$^*$) error grows quickly, while the IMSRG(2$^*$)[3] stays accurate for larger values of $g$, reflecting the approximate cancellation of the missing fourth order quadruples and triples corrections as observed in refs.~\cite{Morris2016,Hergert2016PS}.
%\srs{Say more here.}

%%%%%%%%%%%%%%%%%%%%%%%%%%%%%%%%%%%%%%%
\subsection{Lipkin-Meshkov-Glick model}
%%%%%%%%%%%%%%%%%%%%%%%%%%%%%%%%%%%%%%%

Another toy problem which can illuminate many-body approximation schemes is the Lipkin-Meshkov-Glick (LGM) model~\cite{Lipkin1965}.
The model consists of $N$ particles distributed across two $N$-fold degenerate levels separated by energy $\epsilon$.
The Hamiltonian is
\begin{equation} \label{eq:HLipkin}
    H = \tfrac{1}{2}\epsilon\sum_{p\sigma} \sigma a^{\dagger}_{p\sigma}a_{p\sigma}
    + \tfrac{1}{2} V\sum_{pp'\sigma} a^{\dagger}_{p\sigma}a^{\dagger}_{p'\sigma}a_{p'-\sigma}a_{p-\sigma},
\end{equation}
where $\sigma\in \{-1,1\}$ labels the lower and upper levels, and $p,p'$ run from 1 to $N$, labeling the different degenerate states within a level.

This model can be solved analytically with a quasi-spin formulation, introducing the operators
\begin{equation}
    K_{\pm} = \sum_{p} a^{\dagger}_{p \pm 1}a_{p \mp1}
    \hspace{1em},\hspace{1em}
    K_{z} = \tfrac{1}{2}\sum_{p\sigma}\sigma a^{\dagger}_{p\sigma}a_{p\sigma}
\end{equation}
which obey the usual angular momentum commutation rules (see Appendix~\ref{app:Lipkin}).
%\begin{equation}
%    [K_z,K_{\pm}] = \pm K_{\pm}
%    \hspace{1em},\hspace{1em}
%    [K_+,K_-] = 2K_z.
%\end{equation}
In terms of the quasi-spin operators, the Hamiltonian~\eqref{eq:HLipkin} can be expressed as
\begin{equation}
    H = \epsilon K_z + \tfrac{1}{2} V(K_+^2 + K_-^2).
\end{equation}
$H$ commutes with the operator $K^2=K_z^2 + \tfrac{1}{2}(K_+K_-+K_-K_+)$, which has eigenvalues $k(k+1)$, with $k\leq\frac{N}{2}$.
In the thermodynamic ($N\to\infty$) limit, the LGM model exhibits a phase transition from an ungapped ferromagnetic phase for $NV/\epsilon<1$ to a gapped unmagnetized phase for $NV/\epsilon>1$.

We focus here on the specific case $N=8$, which has the analytical solution\footnote{There appears to be a typo in the solution presented in Ref.~\cite{Lipkin1965}.} for the ground state energy~\cite{Lipkin1965}
\begin{equation}\label{eq:lipkinanalytic}
    \frac{E}{\epsilon} = -\sqrt{10 + 118\left(\frac{V}{\epsilon}\right)^2
    + 6\sqrt{1-2\left(\frac{V}{\epsilon}\right)^2+225\left(\frac{V}{\epsilon}\right)^4}
    }.
\end{equation}

The LGM model has been solved in coupled cluster theory, where it has an analytical result in the doubles (CCD) approximation~\cite{Luhrmann1977} (see also~\cite{Harsha2018}).
The CCD correlation energy is
\begin{equation} \label{eq:ECCD}
    \frac{E_{\rm corr}}{\epsilon} = -\frac{N(N-1)}{2\alpha}\left(1-\sqrt{1-\alpha \frac{V^2}{\epsilon^2}}\right)
\end{equation}
where $\alpha\equiv N^2-7N+9$.
We can see that for $N\geq 6$, $\alpha>0$ and the energy becomes complex for sufficiently large values of $V/\epsilon$.
For the specific case $N=8$ which we consider here, CCD fails for $NV/\epsilon > \sqrt{64/17}\approx 1.94$.
For comparison with the IMSRG(2) we also consider an approximation of CCD in which we omit the intermediate three-body operator arising in the CCD decoupling condition~\cite{Evangelista2012}.
We obtain the same expression for the correlation energy with the modification
$\alpha\to \tilde{\alpha}\equiv N^2-5N+7$.
Note that because this model exhibits an ``excitation-parity'' symmetry, single and triple excitations have no effect, so that CCD is equivalent to CCSDT.

%Moving on to the IMSRG solution, we note that the LGM model was previously considered in a flow equation approach by Pirner and Friman~\cite{Pirner1998}, who used the quasi-spin operators (without normal ordering) to formulate the flow equation (see also Ref.~\cite{Scholtz2003}).
Moving on to the IMSRG solution, we note that the LGM model has been frequently studied within flow equation approach~\cite{Pirner1998,Mielke1998,Kriel2005,Kriel2007,Scholtz2003,Dusuel2005}, notably by Pirner and Friman~\cite{Pirner1998}, who used the quasi-spin operators (without normal ordering) to formulate the flow equation.
They found it necessary to include an induced three-body term, but even then obtained results which were less accurate than second order perturbation theory.
More recently, the flow equation approach was applied to obtain $1/N$ corrections to the thermodynamic limit~\cite{Dusuel2005}.
Here we are interested in finite system, and in understanding the impact of truncations in the particle-rank of the flowing operators.
Making an approximation equivalent to IMSRG(2) requires the use of normal-ordered operators.
Commutators of normal-ordered strings of quasi-spin operators are presented in Appendix~\ref{app:Lipkin}.
While we may construct the IMSRG flow equations in terms of the normal-ordered quasi-spin operators, integrating these equations must still be done numerically.
However, in the Magnus formulation, we may directly solve for the Magnus operator at the fixed point, obtaining (see Appendix~\ref{app:Lipkin})
\begin{equation}
 \tanh\left( \sqrt{ 4\Omega_{\rm 2b}^2 \tilde{\alpha}} \right) = \sqrt{\tilde{\alpha}}\frac{V}{\epsilon}
\end{equation}
and an energy which is identical to the CCD energy with $\alpha\to\tilde{\alpha}$.
Clearly the IMSRG(2) will fail if $\sqrt{\tilde{\alpha}}V/\epsilon>1$.

We now turn to the effects of IMSRG(3).
Due to the excitation-parity symmetry of the LGM model, the three-body pieces of $\eta$ and $\Omega$ vanish.
Consequently, the perturbative triples correction described in section~\ref{sec:perturbative-triples} has no effect.
The only way in which induced three-body operators affect the ground state energy by modifying the flow for $V(s)$ which then subsequently gets contracted with $\Omega$ to generate a contribution to the zero-body term.

The factor of $\tilde{\alpha}$ arises because the double commutator $[K_\pm^2,[K_+^2,K_-^2]]=\pm 8\tilde{\alpha}K_\pm^2$ in the NO2B approximation.
If we also include in our set of flowing operators a term proportional to the three-body operator $\{K_+K_zK_-\}$, we capture the goose-tank diagrams of Fig.~\ref{fig:goosetank}, and the double-commutator is modified so that $\tilde{\alpha}\to\alpha$,  exactly reproducing the CCD energy.
In the IMSRG(2$^*$) approximation, we modify the double commutator by adding the commutator $[\Omega,\chi]$ with the auxiliary operator $\chi$, which also has the effect of replacing $\tilde{\alpha}\to \alpha$, without the need to explicitly construct a three-body operator.

\begin{figure}
    \centering
     \includegraphics[width=\columnwidth]{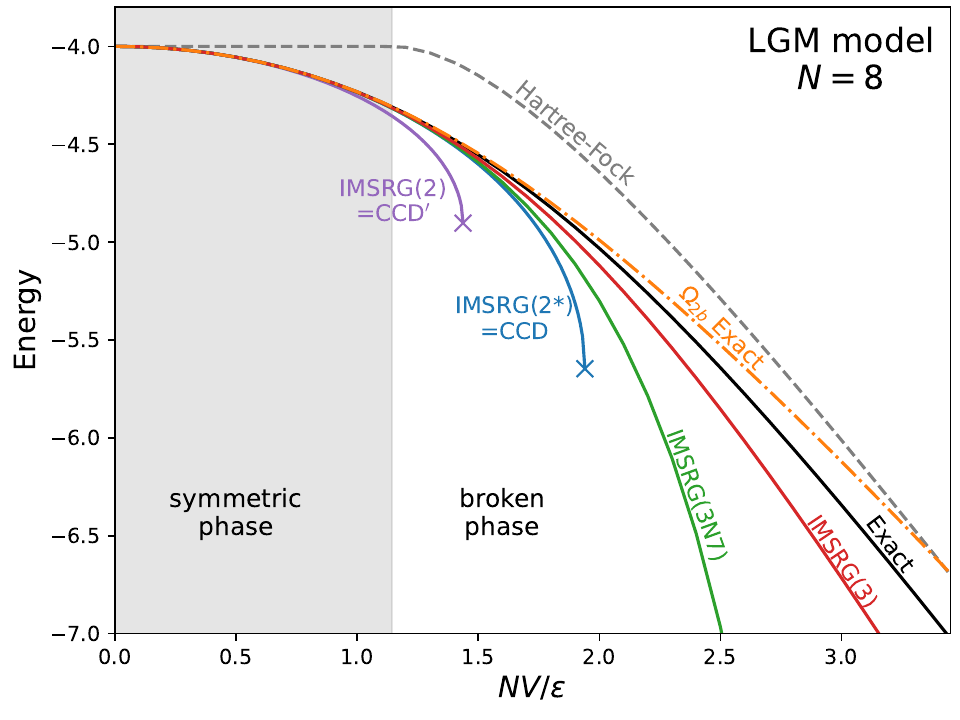}
    \caption{Energy for the $N=8$ LGM model with various approximate methods. See text for details.}
    \label{fig:lipkin}
\end{figure}

There are two additional Hermitian three-body operators consistent with the symmetries of the problem: $\{K_z^3\}$ and $(\{K_+^2K_z\} + \{K_zK_-^2\})$.
Including these spoils the simple structure obtained in the IMSRG(2) approximation, so an analytic solution is less straightforward.
However, it is easy to include them numerically.
The results are shown in Fig.~\ref{fig:lipkin}.

As mentioned above, the IMSRG(2*) energy is identical to the CCD energy, while IMSRG(2) is identical to CCD with the intermediate three-body operator removed, indicated CCD$^\prime$ in Fig.~\ref{fig:lipkin}.
IMSRG(3N7) does better than both of these, and IMSRG(3) gives a significant further improvement.
As mentioned above, due to the excitation-parity symmetry of the problem, there is no improvement from CCSD to CCSDT.
Even in the IMSRG(3) approximation the generator $\eta$ and the Magnus operator $\Omega$ are purely two-body, which means that, as in CCD, both the  IMSRG(2) and IMSRG(3) transformations are specified by a single parameter, $\Omega_{\rm 2b}$.
All the improvement from IMSRG(2) to IMSRG(3) comes from more accurately evaluating the transformation~\eqref{eq:magnusH} by including intermediate three-body operators which subsequently get contracted back down to lower-body operators.
In the specific case in which the three-body operator is immediately contracted back down, e.g. $[\Omega,[\Omega,H]_{\rm 3b}]_{\rm 2b}$, it is possible to factorize the expression for the double nested commutator so that it scales as $N^6$~\cite{Neuscamman2009,Morris2016,Sun2018}.
For the LGM model, including this factorizable correction for all nested commutators in ~\eqref{eq:magnusH} exactly reproduces the IMSRG(3N7) results.
An application to the more realistic case will be presented in a separate work~\cite{He2024}.

Thanks to the simplicity of the LGM model, we may also evaluate the transformation~\eqref{eq:magnusH} directly in the $A$-body space using the quasispin basis.
For $N=8$ this involves the manipulation of $5\times 5$ matrices.
Taking the ansatz $\Omega = \tfrac{1}{2}\Omega_{2b} (K_+^2-K_-^2)$, we solve the flow equations numerically without truncating any intermediate operators.
The result is labeled ``$\Omega_{2b}$ Exact'' in Fig.~\ref{fig:lipkin}.
It lies above the full exact result, as it satisfies the variational principle.

\begin{figure}[t]
     \includegraphics[width=\columnwidth]{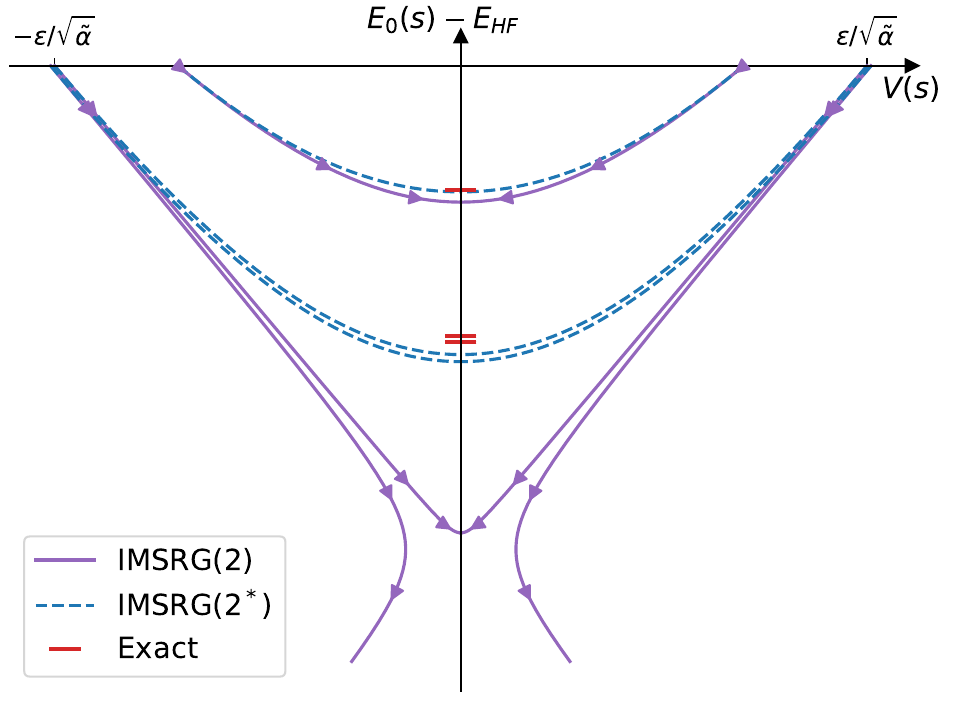}
     \caption{IMSRG flow in the LGM model, for the IMSRG(2) and IMSRG(2$^*$) approximations for several values of $V(0)$.
     Arrows indicate the direction of the flow.
     The red lines indicate the exact solutions from~\eqref{eq:lipkinanalytic}. cf Ref.~\cite{Scholtz2003}.
     \label{fig:LGMflow}}
 \end{figure}

Finally, because the flowing Hamiltonian has essentially two parameters in the IMSRG(2) and IMSRG(2$^*$) approximations, we can visualize the SRG trajectory, as shown in Fig.~\ref{fig:LGMflow}.
(The symmetry about the $x$ axis reflects the symmetry of the LGM model under $V\to -V$.)
Well below the critical coupling strength $\sqrt{\tilde{\alpha}}V/\epsilon=1$, the IMSRG(2) and IMSRG(2$^*$) are both accurate and follow essentially the same trajectory.
Near the critical coupling, the IMSRG(2) diverges significantly from the IMSRG(2$^*$) trajectory, and beyond the critical coupling the fixed point at $V=0$ disappears, so the flow continues to the fixed point at infinity.

To be clear, the observations of this section do not mean that the three-body part of the Magnus operator is unimportant in realistic applications; it is identically zero here because of a symmetry of the LGM model which is not present in general\footnote{Though, for closed harmonic oscillator shells, the usual parity symmetry does impart an approximate excitation-parity symmetry to low-lying states.}.
However, it does demonstrate the significant improvements that can be obtained in the IMSRG by restoring many-body terms inside nested commutators.

\subsection{Realistic nuclear interaction with the valence-space IMSRG}

\begin{figure}[th]
    \centering
        \includegraphics[width=\columnwidth]{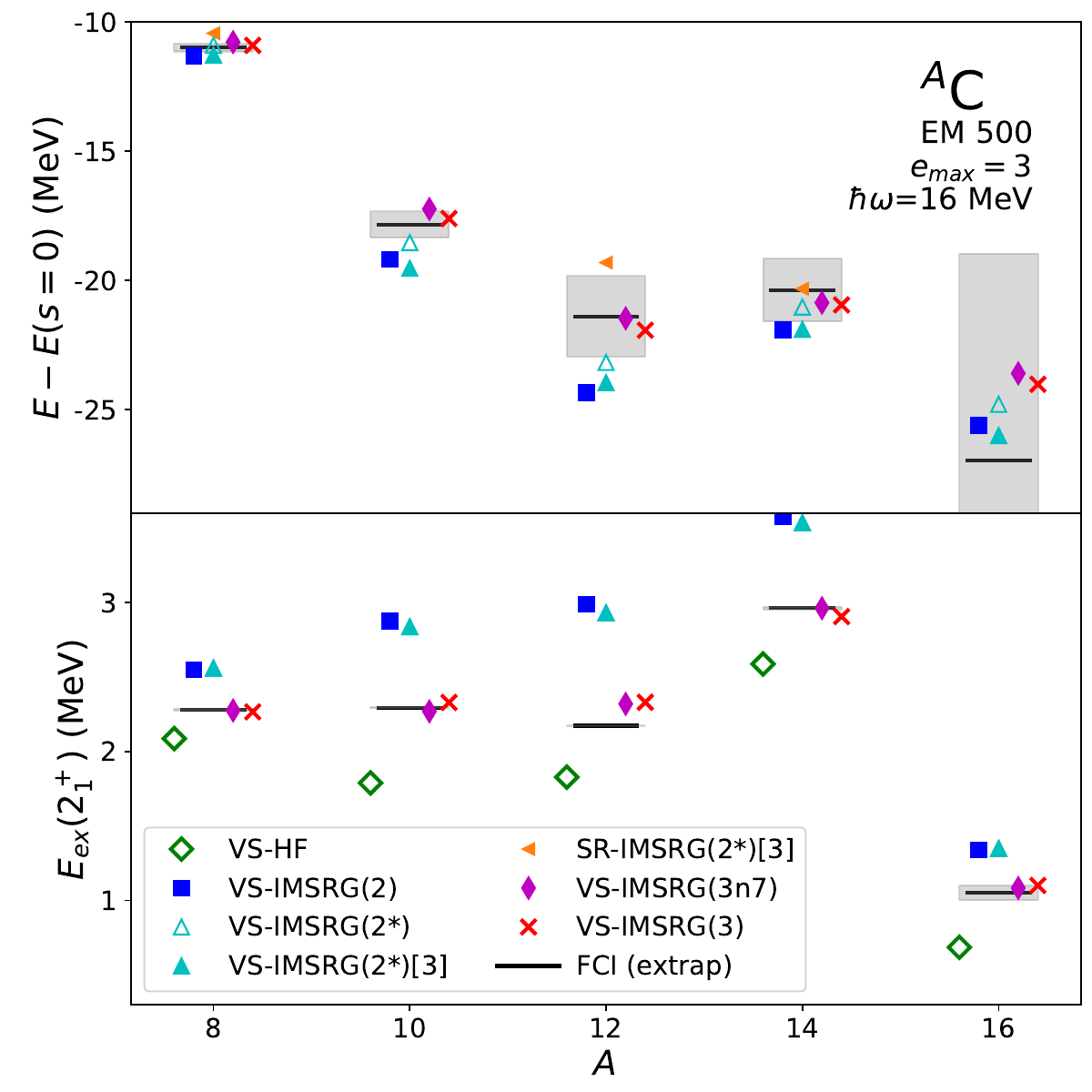}
    \caption{Ground state energy and excitation energy of the first $2^+$ state in carbon isotopes, computed in an oscillator space with $e_{max}=3$, $\hbar\omega=16$~MeV.}
    \label{fig:Cchainbenchmark}
\end{figure}

We next turn to a more realistic nuclear structure calculation, in which we consider the even-$A$ carbon isotopes with $A=8-16$ computed with a pure two-body interaction derived from chiral EFT with a cutoff $\Lambda=500$~MeV and no free-space SRG softening.
We use this relatively hard interaction to emphasize non-perturbative effects.
Our current implementation of the IMSRG(3) limits us to $e_{\rm max}\lesssim 4$, so we cannot perform fully-converged calculations to compare with experiment.
(In any case, comparing with experiment would conflate many-body errors with errors due to the input nuclear force.)
Instead, we compare with truncated configuration interaction (CI) in a given $e_{\rm max}$ space.
This introduces a contamination due to spurious excitations of the center of mass~\cite{Hagen2009,Djarv2021}, but we find that for this specific system the contamination is negligible by $e_{\rm max}=3$.

Ground state energies and $2^+_1$ excitation energies of some even-$A$ carbon isotopes are shown in Fig.~\ref{fig:Cchainbenchmark} for $e_{\rm max}=3$.
To focus on the effect of the IMSRG evolution, for the ground state energies we subtract off the energy obtained by diagonalizing in the valence space with the unevolved $s=0$ Hamiltonian.
Various levels of approximation, from no IMSRG evolution (labeled ``VS-HF'', since we use a Hartree-Fock basis), to full IMSRG(3), are compared with the truncated configuration-interaction (CI) calculations extrapolated to the full CI (FCI) result.
We truncate on $N_{\rm max}$, i.e. the maximum number of harmonic oscillator quanta of excitation, up to $N_{\rm max}=$~6 or 8.
The gray error bands in Fig.~\ref{fig:Cchainbenchmark} indicate the difference from the largest $N_{\rm max}$ calculation performed to the $N_{\rm max}\to \infty$ extrapolated value.
The first observation is that going from IMSRG(2) to IMSRG(3) generally improves agreement with FCI for both the ground state energy and the excitation energy.
We also notice that the IMSRG(3N7) approximation accurately reproduces the full IMSRG(3) result.
The IMSRG(2$^*$) approximation overbinds in the valence space formulation and slightly underbinds in the single-reference formulation.
Nevertheless, it is notable that the SR-IMSRG(2$^*$) flow converges for $^{12}$C, while it diverges for SR-IMSRG(2).

The IMSRG(2) approximation consistently yields excitation energies that are too high, while this is corrected by going to IMSRG(3N7). 
It appears that the most important modification to the effective valence space interaction in going from IMSRG(2) to IMSRG(3N7) is that the off-diagonal matrix element $\langle p_{3/2} p_{3/2}|V|p_{1/2} p_{1/2}\rangle_{J=0}$ is reduced in magnitude by $\sim \! 0.5$~MeV.
This term, which is too large in the IMSRG(2) approximation, drives 2p-2h correlations which provide too much correlation energy for the ground state.
These correlations are less active in the $2^+$ excited state, and so the net result is an over-predicted excitation energy.
It is also interesting to note that the IMSRG(2*) approximation has essentially no impact on these off-diagonal valence space matrix elements, and thus gives no improvement for the $2^+$ excitation energy.
This can be understood because the IMSRG(2*) approximation essentially generates the goose-tank diagrams of Fig.~\ref{fig:goosetank} via a contraction of a one-body operator $\chi$ with the two-body part of the generator, so that the main effect is to modify the off-diagonal part of $H$.
Valence-to-valence matrix elements are not included in the definition of off-diagonal and so this term receives relatively little modification.

\section{\label{sec:cost}Computational cost}
\begin{figure}
    \centering
    \includegraphics[width=\columnwidth]{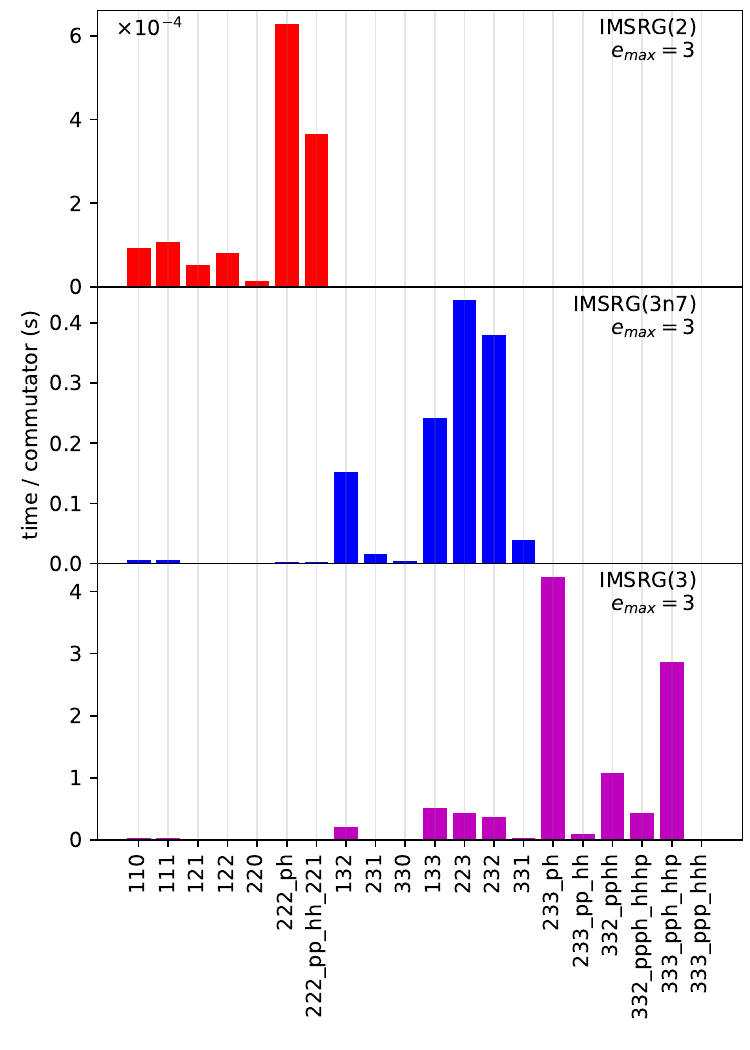}
    \caption{\label{fig:timingbyterm} Compute time per commutator on a single node with 64 threads, for three levels of approximation. The labels indicate the particle rank of the two operators entering the expression and the resulting operator. The following strings of p's and h's indicate whether the contracted indices are particle or hole, for cases where the routine was split up according to this feature.}
\end{figure}

Finally, we briefly discuss the computational cost of our current implemententation of the IMSRG(3), and approximations.
While this is to some extent specific to this implementation, other implementations will share the general features, and this also helps to illuminate where future improvements should be focused.
All commutator expressions involving a three-body operator have been recast as matrix matrix multiplications, leading to significant speedup with the exception of the $[2,2]\to 3$ expression.
In this case, we decided that the modest speedup was not worth the considerable complication in the code.
Multi-threaded parallelism via OpenMP is utilized throughout these routines.
The time for each expression is shown in Fig.~\ref{fig:timingbyterm}.
Which commutator expression is the most expensive depends on the model space and system studied, but beyond the smallest spaces the $[2,2]\to 3$ and $[2,3]\to 2$ expressions are the main bottleneck of IMSRG(3N7) calculations.
For full IMSRG(3), terms with intermediate pph lines dominate, due to the angular momentum recoupling required to cast them as matrix multiplications.
We observe that the naive scaling estimate is indeed naive; the $[3,3]\to 3$ commutator with ppp or hhh intermediate states scales as $N^9$, but it takes a negligible amount of time because it is highly amenable to matrix multiplication without any required recouplings or antisymmetrizations.

\section{Conclusion}

We have performed a detailed exploration of the impact of retaining three-body operators within the IMSRG, building on previous work~\cite{Heinz2021}.
While a full inclusion of three-body operators remains too expensive for practical calculations, we find that more manageable approximations capture the most important corrections.
We explored two toy models, selected to emphasize short-range and long-range correlations, respectively.
We find that in the case of a short-range interaction, the main error made in the IMSRG(2) is corrected by including the effects of the goose-tank diagrams in Fig.~\ref{fig:goosetank}, which can be done without changing the computational scaling.
We also investigated the origin of the diverging IMSRG flow for sufficiently large coupling constant $g$, and related it to coherent effects due to the approximately separable nature of the interaction; these coherent effects were offset by including the missing goose-tank diagrams, stabilizing the flow.

In the case of long-range correlations, as captured by the LGM model, we presented an analytic solution for the IMSRG(2) approximation, and showed that including intermediate three-body can lead to a significant improvement in accuracy over coupled cluster.
Finally, we have explored the impact of three-body operators in the valence-space IMSRG with a realistic interaction.
We found that the systematic over-prediction of $2^+$ excitation energies in the IMSRG(2) approximation is essentially corrected by including three-body operators during the IMSRG evolution, and that this improvement is achieved already at the IMSRG(3N7) level of approximation.
Finally, we have presented arguments, supported by our calculations, that the most important effect of three-body operators arises as intermediate terms in nested commutators.
The leading corrections of this type can be factorized to scale as $N^6$, and so may be employed in large-scale calculations.
This is presented in a separate paper~\cite{He2024}.

\begin{acknowledgments}
S.R.S. would like to thank Matthias Heinz for help on optimizing the implementation and finding a typo in the commutator expressions.
S.R.S. and B.C.H. acknowledge support from the U.S. Department of Energy (DoE), Office of Science, under SciDAC-5 (NUCLEI collaboration) under grant DE-FG02-97ER41014.
This research was supported in part by the Notre Dame’s Center for Research Computing.
This work benefitted from the use of the \texttt{amc} code~\cite{Tichai2020} to validate the $J$-coupled expressions.
The \texttt{imsrg++} code~\cite{imsrg++} employed in this work uses the \texttt{Armadillo} library~\cite{Armadillo2016,Armadillo2019}.
\end{acknowledgments}

\appendix

\section{Uncoupled commutator expressions\label{app:commutators_m}}
We need the expression for the commutator $C=[A,B]$ where $A$, $B$, and $C$ are all Fock-space operators including up to 3 body operators, e.g.
\begin{equation}
\begin{aligned}
    C = C_0& + \sum_{ij}C_{ij} \{a^{\dagger}_ia_j\} + \frac{1}{4}\sum_{ijkl}C_{ijkl}\{a^{\dagger}_ia^{\dagger}_ja_la_k\} \\
    &+\frac{1}{36}\sum_{ijklmn}C_{ijklmn}\{a^{\dagger}_ia^{\dagger}_ja^{\dagger}_ka_na_ma_l\}
    \end{aligned}
\end{equation}
and likewise for $A$ and $B$.
In the following, the permutation operator $P_{ij}$ exchanges the indices $i$ and $j$ of the expression to its right, with no implicit minus sign. For example $P_{ij}A_{bjal}B_{aibk}=A_{bial}B_{ajbk}$.
We also use the operator
\begin{equation}
    P_{ij/k} \equiv 1 - P_{ik} - P_{jk}.
\end{equation}
\begin{widetext}
\begin{equation}\label{eq:mschemeC0}
\begin{aligned}
    C_0 = &\sum_{ab}(n_a-n_b)A_{ab}B_{ba}
    + \frac{1}{4}\sum_{abcd}n_an_b\bar{n}_c\bar{n}_d (A_{abcd}B_{cdab}-B_{abcd}A_{cdab})\\
    &+ \frac{1}{36}\sum_{abcdef}n_an_bn_c\bar{n}_d\bar{n}_e\bar{n}_f
    (A_{abcdef}B_{defabc}-B_{abcdef}A_{defabc})
    \end{aligned}
\end{equation}
\begin{equation}\label{eq:mschemeC1}
\begin{aligned}
 C_{ij} = &\sum_{a}(A_{ia}B_{aj}-B_{ia}A_{aj})
 + \sum_{ab}(n_a-n_b)(A_{ab}B_{biaj}-B_{ab}A_{biaj})\\
 &+\frac{1}{2}\sum_{abc}(n_an_b\bar{n}_c+\bar{n}_a\bar{n}_bn_c)
 (A_{ciab}B_{abcj}-B_{ciab}A_{abcj}) \\
 &+\frac{1}{4}\sum_{abcd}n_an_b\bar{n}_c\bar{n}_d(A_{abcd}B_{cdiabj} +A_{abicdj}B_{cdab} -B_{abcd}A_{cdiabj}-B_{abicdj}A_{cdab})\\
 &+\frac{1}{12}\sum_{abcde}(n_an_b\bar{n}_c\bar{n}_d\bar{n}_e + \bar{n}_a\bar{n}_bn_cn_dn_e)(A_{abicde}B_{cdeabj}-B_{abicde}A_{cdeabj})
\end{aligned}
\end{equation}
\begin{equation}\label{eq:mschemeC2}
\begin{aligned}
    C_{ijkl} = &\sum_{a}(A_{ia}B_{ajkl}+A_{ja}B_{iakl}-B_{ijal}A_{ik}-B_{ijka}A_{al})\\
    &-\sum_{a}( B_{ia}A_{ajkl}+B_{ja}A_{iakl}-A_{ijal}B_{ka}-A_{ijka}B_{ak}) \\
    &+\frac{1}{2}\sum_{ab}(\bar{n}_a\bar{n}_b-n_an_b)(A_{ijab}B_{abkl}-B_{ijab}A_{abkl})\\
    &-\sum_{ab}(n_a-n_b)(1-P_{ij})(1-P_{kl})A_{bjal}B_{aibk}\\
    &+\sum_{ab}(n_a-n_b) ( A_{ab}B_{ijbkla} - B_{ab}A_{ijbkla})\\
%    &+\frac{1}{2}\sum_{abc}(n_an_b\bar{n}c+\bar{n}_a\bar{n}_bn_c) (A_{icab}B_{abjklc}+A_{cjab}B_{abiklc}-B_{ijcabl}A_{abkc}-B_{ijcabk}A_{abcl})\\
    &+\frac{1}{2}\sum_{abc}(n_an_b\bar{n}_c+\bar{n}_a\bar{n}_bn_c)
    (1-P_{ij})(A_{icab}B_{ajbklc}-B_{icab}A_{abjklc})\\
    &-\frac{1}{2}\sum_{abc}(n_an_b\bar{n}_c+\bar{n}_a\bar{n}_bn_c)
    (1-P_{kl})(B_{ijcabl}A_{abkc}-A_{ijcabl}B_{abkc})\\
%    (B_{icab}A_{abjklc}+B_{cjab}A_{abiklc}-A_{ijcabl}B_{abkc}-A_{ijcabk}B_{abcl})\\
    &+\frac{1}{6}\sum_{abcd}(n_an_bn_c\bar{n}_d-\bar{n}_a\bar{n}_b\bar{n}_cn_d)(A_{ijdabc}B_{abckld} - B_{ijdabc}A_{abckld})\\
    &+\frac{1}{4}\sum_{abcd}(\bar{n}_a\bar{n}_bn_cn_d-n_an_b\bar{n}_c\bar{n}_d)(1-P_{ij})(1-P_{kl})A_{abicdk}B_{cdjabl}
    \end{aligned}
\end{equation}
\begin{equation}\label{eq:mschemeC3}
\begin{aligned}
    C_{ijklmn} = &\sum_{a}P_{i/jk}(A_{ia}B_{ajklmn}-B_{ia}A_{ajklmn}) -P_{l/mn}(B_{ijkamn}A_{al}-A_{ijkamn}B_{al})\\
%    &+\sum_{a}P_{ij/k}P_{lm/n}(A_{ijla}B_{akmn}-B_{ijla}A_{akmn})\\
    &-\sum_{a}P_{ij/k}P_{lm/n}(A_{ijna}B_{kalm}-B_{ijna}A_{kalm})\\
    &+\frac{1}{2}\sum_{ab}(\bar{n}_a\bar{n}_b-n_an_b)
     P_{ij/k}(A_{ijab}B_{abklmn}-B_{ijab}A_{abklmn})\\
     &-\frac{1}{2}\sum_{ab} (\bar{n}_a\bar{n}_b-n_an_b) P_{lm/n}(B_{ijkabn}A_{ablm}-A_{ijkabn}B_{ablm})\\
     &-\sum_{ab}(n_a \bar{n}_b -\bar{n}_a n_b)P_{ij/k}P_{lm/n} (A_{bkan} B_{ijalmb}-B_{bkan}A_{ijalmb})\\
     &+\frac{1}{6}\sum_{abc}(n_an_bn_c+\bar{n}_a\bar{n}_b\bar{n}_c)(A_{ijkabc}B_{abclmn}-B_{ijkabc}A_{abclmn})\\
     &-\frac{1}{2}\sum_{abc}(n_an_b\bar{n}_c+\bar{n}_a\bar{n}_bn_c)P_{ij/k}P_{lm/n}(A_{ijcabn}B_{abklmc}-B_{ijcabn}A_{abklmc})
     \end{aligned}
\end{equation}
\end{widetext}

\section{Commutator expressions in \texorpdfstring{$J$}{J}-coupled form\label{app:commutators_j}}
To take advantage of rotational symmetry, we work with $J$-coupled matrix elements.
To do this, we switch our notation so that we now explicitly indicate the projection quantum number $m$ in the uncoupled basis.
One body terms are written as
\begin{equation}\label{eq:Coupled1body}
    C_{im_ijm_j} = C_{ij}\delta_{m_im_j}
\end{equation}
while the two and three-body terms are written in terms of unnormalized $J$-coupled matrix elements
%\begin{widetext}
\begin{equation}\label{eq:Coupled2body}
%    C_{im_ijm_jkm_klm_l} = \sum_{JM}\langle im_ijm_j|JM\rangle\langle km_klm_l|JM\rangle C^{J}_{ijkl}
    C_{im_ijm_jkm_klm_l} = \sum_{JM}\mathcal{C}_{j_im_ij_jm_j}^{JM} \mathcal{C}_{j_km_kj_lm_l}^{JM}  C^{J}_{ijkl}
\end{equation}
\begin{equation}\label{eq:Coupled3body}
\begin{aligned}
    C_{im_i jm_j km_k lm_l m m_m n m_n}=\sum_{\substack{J_1J_2\mathcal{J}\\M_1M_1\mathcal{M}}} 
    \mathcal{C}_{j_im_ij_jm_j}^{J_1M_1} \mathcal{C}_{J_1M_1 j_km_k}^{\mathcal{J}\mathcal{M}} &  \\
    \times
    \mathcal{C}_{j_lm_lj_mm_m}^{J_2M_2} \mathcal{C}_{J_2M_2 j_nm_n}^{\mathcal{J}\mathcal{M}} 
%    \langle im_ijm_j | J_1M_1\rangle \langle J_1M_1 km_k|\mathcal{J}\mathcal{M}\rangle      \langle lm_lmm_m | J_2M_2\rangle\langle J_2M_2 nm_n | \mathcal{J}\mathcal{M}\rangle
    C^{J_1J_2\mathcal{J}}_{ijklmn}.&
    \end{aligned}
\end{equation}
%\end{widetext}
The $\mathcal{C}$ in \eqref{eq:Coupled2body} and \eqref{eq:Coupled3body} are Clebsch-Gordan coefficients.
By inserting \eqref{eq:Coupled1body} \eqref{eq:Coupled2body} and \eqref{eq:Coupled3body} into \eqref{eq:mschemeC0}, \eqref{eq:mschemeC1}, \eqref{eq:mschemeC2}, and \eqref{eq:mschemeC3}, we obtain the $J$-coupled commutator expressions.

In order to make the expressions somewhat more compact, in the following we use orbit labels to indicate the associated angular momentum, e.g. in 6$j$, 9$j$ symbols and phase factors so that $(-1)^{i+j-J}$ is shorthand for $(-1)^{j_j+j_j-J}$.
We also introduce the coupled permutation operators
\begin{equation}
    P^{J}_{ij} \equiv (-1)^{i+j-J}P_{ij}
\end{equation}
\begin{equation}\label{eq:CoupledPijk}
\begin{aligned}
    P^{J_1\mathcal{J}}_{ij/k} \equiv 1 &- \sum_{J_1'}\langle ijk J_1\mathcal{J} | kji J_1'\mathcal{J}\rangle P_{ik}^{J_1\rightarrow J_1'}\\
    &-\sum_{J_1'}\langle ijk J_1\mathcal{J} | ikj J_1'\mathcal{J}\rangle P_{jk}^{J_1\rightarrow J_1'}
    \end{aligned}
\end{equation}
The brackets in \eqref{eq:CoupledPijk} are recoupling coefficients which may be expressed in terms of 6$j$ symbols
\begin{equation}
    \langle ijk J_1\mathcal{J} | kji J_1'\mathcal{J}\rangle = 
    \hat{J}_1\hat{J}_1'(-1)^{2i+2j+2k}
    \begin{Bmatrix}
    i & j & J_1 \\ k & \mathcal{J} & J_1'
    \end{Bmatrix}
\end{equation}
and 
\begin{equation}
    \langle ijk J_1\mathcal{J} | ikj J_1'\mathcal{J}\rangle = 
    \hat{J}_1\hat{J}_1'(-1)^{j+k+J_1+J_1'}
    \begin{Bmatrix}
    j & i & J_1 \\ k & \mathcal{J} & J_1'
    \end{Bmatrix}.
\end{equation}
The notation $J_1\rightarrow J_1'$ in the superscript in \eqref{eq:CoupledPijk} indicates that every instance of $J_1$ to the right of the permutation operator should be replaced with $J_{1}'$.
We use the usual notation $\hat{J}\equiv \sqrt{2J+1}$.
With this notation, we have for our antisymmetrized matrix elements
\begin{equation}
    P^{J}_{ij}C^{J}_{ijkl} = -C^{J}_{ijkl}
\end{equation}
and
\begin{equation}
    P^{J_1\mathcal{J}}_{ij/k} C^{J_1J_2\mathcal{J}}_{ijklmn} = 3~C^{J_1J_2\mathcal{J}}_{ijklmn}.
\end{equation}

As in previous presentations, we employ Pandya-transformed operators, denoted with an overbar
\begin{equation}
    \bar{C}^{J}_{i\bar{j}k\bar{l}} \equiv -\sum_{J'}\hat{J'}^2
    \begin{Bmatrix}
    i & j & J \\
    k & l & J'
    \end{Bmatrix}
    C^{J'}_{ilkj}.
\end{equation}
We also define a transformed three-body operator as
\begin{equation}
\begin{aligned}
    \bar{C}^{J'}_{i\bar{l};(cdJ_{cd})(\overline{abJ_{ab}})} \equiv -\sum_{\mathcal{J}} &\hat{\mathcal{J}}^2(-1)^{i+\mathcal{J}}
    \begin{Bmatrix}
    i & l & J'\\ J_{cd} & J_{ab} & \mathcal{J}
    \end{Bmatrix} \\
  &\times  C^{J_{ab}J_{cd}\mathcal{J}}_{abicdl}.
    \end{aligned}
\end{equation}
With these definitions, the $J$-coupled commutator expressions become
\begin{widetext}
\begin{equation}
\begin{aligned}
    C_{0} = &\sum_{ab}\hat{j}^2_a (n_a-n_b) A_{ab}B_{ba}
    +\frac{1}{4}\sum_{abcd}\sum_{J}\hat{J}^2 n_an_b\bar{n}_c\bar{n}_d(A^{J}_{abcd}B^{J}_{cdab}-B^{J}_{abcd}A^{J}_{cdab})\\
    &+\frac{1}{36}\sum_{abcdef}\sum_{J_1J_2\mathcal{J}}n_an_bn_c\bar{n}_d\bar{n}_e\bar{n}_f \hat{\mathcal{J}}^2 
    (A^{J_1J_2\mathcal{J}}_{abcdef} B^{J_2J_1\mathcal{J}}_{defabc}
    -B^{J_1J_2\mathcal{J}}_{abcdef} A^{J_2J_1\mathcal{J}}_{defabc})
    \end{aligned}
\end{equation}
\begin{equation}
\begin{aligned}
    C_{ij} = &\sum_{a}(A_{ia}B_{aj} - B_{ia}A_{aj}) % typo found by Bingcheng
    +\sum_{ab}\sum_{J}(n_a-n_b)\frac{\hat{J}^2}{\hat{j}_i^2}(A_{ab}B^{J}_{biaj}-B_{ab}A^{J}_{biaj})\\
    &+ \frac{1}{2}\sum_{abc}\sum_{J}(n_an_b\bar{n}_c+\bar{n}_a\bar{n}_bn_c)\frac{\hat{J}^2}{\hat{j}_i^2}
    (A^{J}_{ciab}B^{J}_{abcj}
    -B^{J}_{ciab}A^{J}_{abcj})\\
    &+\frac{1}{4}\sum_{abcd}\sum_{J\mathcal{J}}n_an_b\bar{n}_c\bar{n}_d \frac{\hat{\mathcal{J}}^2}{\hat{j}_i^2} 
    (A^{J}_{abcd}B^{JJ\mathcal{J}}_{cdiabj}
    +A^{JJ\mathcal{J}}_{abicdj}B^{J}_{cdab}
    -B^{J}_{abcd}A^{JJ\mathcal{J}}_{cdiabj}
    -B^{JJ\mathcal{J}}_{abicdj}A^{J}_{cdab})\\
    &+\frac{1}{12}\sum_{abcde}\sum_{J_1J_2\mathcal{J}}(n_an_b\bar{n}_c\bar{n}_d\bar{n}_e +\bar{n}_a\bar{n}_bn_cn_dn_e)\frac{\hat{\mathcal{J}}^2}{\hat{j}_i^2}
    (A^{J_1J_2\mathcal{J}}_{abicde}B^{J_2J_1\mathcal{J}}_{cdeabj}
    -B^{J_1J_2\mathcal{J}}_{abicde}A^{J_2J_1\mathcal{J}}_{cdeabj})
\end{aligned}
\end{equation}
\begin{equation}
    \begin{aligned}
        C^{J}_{ijkl} =&
        \sum_{a}(A_{ia}B^{J}_{ajkl}+A_{ja}B^{J}_{iakl}
                -B^{J}_{ijal}A_{ak}-B^{J}_{ijka}A_{al})\\
        &-\sum_{a}(B_{ia}A^{J}_{ajkl}+B_{ja}A^{J}_{iakl}
                -A^{J}_{ijal}B_{ak}-A^{J}_{ijka}B_{al})\\
        &+\frac{1}{2}\sum_{ab}\sum_{J}(\bar{n}_a\bar{n}_b-n_an_b)(A^{J}_{ijab}B^{J}_{abkl}-B^{J}_{ijab}A^{J}_{abkl})\\
%%        &-\sum_{ab}\sum_{J'}(n_a-n_b)\hat{J}^{'2} (1-P^{J}_{ij}) %%% Wrong sign on the ph term. Oops.
        &+\sum_{ab}\sum_{J'}(n_a-n_b)\hat{J}^{'2} (1-P^{J}_{ij})
        \begin{Bmatrix}
        i & j & J\\  k & l & J'
        \end{Bmatrix}
        (\bar{A}^{J'}_{i\bar{l}a\bar{b}}\bar{B}^{J'}_{a\bar{b}k\bar{j}}
        -\bar{B}^{J'}_{i\bar{l}a\bar{b}}\bar{A}^{J'}_{a\bar{b}k\bar{j}})\\
        &+\sum_{ab}\sum_{\mathcal{J}}(n_a-n_b)\frac{\hat{\mathcal{J}}^2}{\hat{J}^2} (A_{ab}B^{JJ\mathcal{J}}_{ijbkla} % typo found by Bingcheng
        -B_{ab}A^{JJ\mathcal{J}}_{ijbkla})\\
        &-\frac{1}{2}\sum_{abc}\sum_{J'\mathcal{J}}(n_an_b\bar{n}_c+\bar{n}_a\bar{n}_bn_c)\frac{\hat{J}'\hat{\mathcal{J}}^{2}}{\hat{J}} \Bigl[
        (1-P_{ij}^J)
        \begin{Bmatrix}
        i & j & J \\  c&\mathcal{J} & J'
        \end{Bmatrix}
        (A^{J'}_{cjab}B^{J'J\mathcal{J}}_{abiklc}-B^{J'}_{cjab}A^{J'J\mathcal{J}}_{abiklc})\\
        & \hspace{15em}-
%        &+\frac{1}{2}\sum_{abc}\sum_{J'\mathcal{J}}(n_an_b\bar{n}_c+\bar{n}_a\bar{n}_bn_c)\frac{\hat{J}'\hat{\mathcal{J}}^{2}}{\hat{J}}
        (1-P_{kl}^J)
        \begin{Bmatrix}
        k & l & J \\ c& \mathcal{J} & J'
        \end{Bmatrix}
        (B^{JJ'\mathcal{J}}_{ijcabk}A^{J'}_{abcl}-A^{JJ'\mathcal{J}}_{ijcabk}B^{J'}_{abcl})  \Bigr] \\
        &+\frac{1}{6}\sum_{abcd}\sum_{J'\mathcal{J}}(n_an_bn_c\bar{n}_d-\bar{n}_a\bar{n}_b\bar{n}_cn_d)\frac{\hat{\mathcal{J}}^2}{\hat{J}^2} % typo found by SRS and fixed. The 2nd occ factor needs a minus sign
        (A^{JJ'\mathcal{J}}_{ijdabc}B^{J'J\mathcal{J}}_{abckld}
        -B^{JJ'\mathcal{J}}_{ijdabc}A^{J'J\mathcal{J}}_{abckld})\\
        &+\frac{1}{4}\sum_{\substack{abcd\\J_{ab}J_{cd}}}\sum_{J_{ph}}(\bar{n}_a\bar{n}_bn_cn_d-n_an_b\bar{n}_c\bar{n}_d)
        (1-P^{J}_{ij})(1-P^{J}_{kl})
%        \hat{\mathcal{J}}^2
        \hat{J_{ph}}^2 % typo found by Bingcheng
        \begin{Bmatrix}
        i & j & J\\ k & l & J_{ph} % typo found by Bingcheng
        \end{Bmatrix}
        \bar{A}^{J_{ph}}_{i\bar{l};(cdJ_{cd})(\overline{abJ_{ab}})}
        \bar{B}^{J_{ph}}_{(cdJ_{cd})(\overline{abJ_{ab}});k\bar{j}}
%        &+\frac{1}{4}\sum_{abcd}\sum_{J'J''\mathcal{J}\mathcal{J}'}(\bar{n}_a\bar{n}_bn_cn_d-n_an_b\bar{n}_c\bar{n}_d)
%        \hat{\mathcal{J}}^2\hat{\mathcal{J}}^{'2}
%        (1-P^{J}_{ij})(1-P^{J}_{kl})
%        (-1)^{i+k+\mathcal{J}+\mathcal{J}'}
%        \begin{Bmatrix}
%        J' & k & \mathcal{J}'\\
%        i & J & j \\
%        \mathcal{J} & l & J''
%        \end{Bmatrix}
%        A^{J'J''\mathcal{J}}_{abicdl}B^{J''J'\mathcal{J}'}_{cdjabk}
        \end{aligned}
\end{equation}
\begin{equation}
    \begin{aligned}
        C^{J_1J_2\mathcal{J}}_{ijklmn} =&
    \sum_{a}\left\{ P^{J_1\mathcal{J}}_{ij/k} (A_{ka}B^{J_1J_2\mathcal{J}}_{ijalmn}-B_{ka}A^{J_1J_2\mathcal{J}}_{ijalmn})
    - P^{J_2\mathcal{J}}_{lm/n} (B^{J_1J_2\mathcal{J}}_{ijklma}A_{an}-A^{J_1J_2\mathcal{J}}_{ijklma}B_{an}) \right\}\\
    &+\sum_{a}P^{J_1\mathcal{J}}_{ij/k}P^{J_2\mathcal{J}}_{lm/n}
    \hat{J}_1\hat{J}_2
    \begin{Bmatrix}
    n & a & J_1 \\ k & \mathcal{J} & J_2
    \end{Bmatrix}
    (A^{J_1}_{ijna}B^{J_2}_{kalm}-B^{J_1}_{ijna}A^{J_2}_{kalm})\\
    &+\frac{1}{2}\sum_{ab}(\bar{n}_a\bar{n}_b-n_an_b)
    P^{J_1\mathcal{J}}_{ij/k}
    (A^{J_1}_{ijab}B^{J_1J_2\mathcal{J}}_{abklmn}-B^{J_1}_{ijab}A^{J_1J_2\mathcal{J}}_{abklmn})\\
    &-\frac{1}{2}\sum_{ab}(\bar{n}_a\bar{n}_b-n_an_b)
    P^{J_2\mathcal{J}}_{lm/n}
    (B^{J_1J_2\mathcal{J}}_{ijkabn}A^{J_2}_{ablm}-A^{J_1J_2\mathcal{J}}_{ijkabn}B^{J_2}_{ablm})\\
    &-\sum_{ab}\sum_{J'\mathcal{J}'}(n_a\bar{n}_b- \bar{n}_a n_b)\hat{J}^{'2} \hat{\mathcal{J}}^{'2}
    P^{J_1\mathcal{J}}_{ij/k} P^{J_2\mathcal{J}}_{lm/n}
%    (-1)^{k+n+J_1+J_2}
    (-1)^{k+n+J_1+J_2+\mathcal{J}}
    \begin{Bmatrix}
    b & \mathcal{J}' & J_2\\
    k & J_1 & \mathcal{J} \\
    J' & a & n
    \end{Bmatrix}
    (A^{J'}_{bkan}B^{J_1J_2\mathcal{J}'}_{ijalmb}-B^{J'}_{bkan}A^{J_1J_2\mathcal{J}'}_{ijalmb})\\
    &+\frac{1}{6}\sum_{abc}\sum_{J'}(n_an_bn_c+\bar{n}_a\bar{n}_b\bar{n}_c)
    (A^{J_1J'\mathcal{J}}_{ijkabc}B^{J'J_2\mathcal{J}}_{abclmn}
    -B^{J_1J'\mathcal{J}}_{ijkabc}A^{J'J_2\mathcal{J}}_{abclmn})\\
    &+\frac{1}{2}\sum_{abc}\sum_{J'\mathcal{J}'\mathcal{J}''}
    (n_an_b\bar{n}_c+\bar{n}_a\bar{n}_bn_c)
    \hat{\mathcal{J}}^{'2}\hat{\mathcal{J}}^{''2}
    P^{J_1\mathcal{J}}_{ij/k} P^{J_2\mathcal{J}}_{lm/n}
    \begin{Bmatrix}
    k & J' & \mathcal{J}' \\
    J_1 & \mathcal{J}'' & c \\
    \mathcal{J} & n & J_2
    \end{Bmatrix}
    (A^{J'J_2\mathcal{J}'}_{abklmc}B^{J_1J'\mathcal{J}''}_{ijcabn}
    -B^{J'J_2\mathcal{J}'}_{abklmc}A^{J_1J'\mathcal{J}''}_{ijcabn})
    \end{aligned}
\end{equation}
\end{widetext}

\section{Normal-ordered commutator for the LGM model\label{app:Lipkin}}
Operators built from the usual quasi-spin operators $K_{\pm},K_z$ can be re-expressed in terms of operators normal-ordered with respect to the mean-field ground state in the unbroken phase (we indicate the normal-ordered operators with braces $\{\}$) :
\begin{align}
    K_z &= \{ K_z \} -\tfrac{1}{2}N \\
    K_z^2 &= \{ K_z^2 \} + (\tfrac{1}{2}\!-\!N)\{K_z\} + \tfrac{1}{4}N^2 \\
    K_z^3 &= \{ K_z^3 \} + \tfrac{3}{2}(1-N)\{K_z^2\} \nonumber \\ &~~+ \tfrac{1}{4}(1-3N+3N^2)\{K_z\}-\tfrac{1}{8}N^3 \\
    K_{\pm}^n &= \{K_{\pm}^n \}  \\
%    K_{\pm}^2K_z &= \{K_{\pm}^2 K_z\} -(\tfrac{1}{2}N\pm 1)\{J_{\pm}^2\} \\
    K_{+}^2K_z &= \{K_{+}^2 K_z\} -\tfrac{1}{2}N\{K_{+}^2\} \\
    K_zK_{-}^2 &= \{K_zK_{-}^2\} -\tfrac{1}{2}N\{K_{-}^2\} \\
    K_{+}K_{-} &= \{K_{+}K_{-} \} \\
    K_{-}K_{+} &= \{K_{+}K_{-} \} -2\{K_z\} +N \\
%    K_{+}K_{-}+K_{-}K_{+} &= 2K^2 -2K_z^2\\
%    &=\{K_{+}K_{-} +K_{-}K_{+}\} - 2\{K_z\} +N \nonumber \\
    K_{+}K_{z}K_{-} &= \{K_{+}K_{z}K_{-}\} -\tfrac{1}{2}N \{K_{+}K_{-}\}
\end{align}
%It is also helpful to note
%\begin{equation}
%    \{K_+^m\}\{K_z^n\}\{K_-^l\}=\{K_+^m K_z^n K_-^l\}.
%\end{equation}
Relevant commutator expression in terms of the vacuum normal ordered operators are
\begin{align}\label{eq:LipkinCommNO}
    [K_{z},K_{\pm}] &= \pm K_{\pm} \\
    [K_{+},K_{-}] &= 2K_{z} \\
    [K_+^2,K_-^2] &= 8 K_{+}K_{z}K_{-}+8K_{+}K_{-} \nonumber \\ & ~~ - 8K_z^2 - 4K_z \\
    [K_z,K_{\pm}^2] &= \pm 2 K_{\pm}^2 \\ 
    [K_z^2,K_{+}^2] &=  4 K_{+}^2K_z + 4K_{+}^2 \\
    [K_z^2,K_{-}^2] &= - 4 K_zK_{-}^2 - 4K_{-}^2 \\
    [K_{+}K_{-},K_{+}^2] &= -4K_{+}^2K_z - 2K_{+}^2 \\
    [K_{+}K_{-},K_{-}^2] &= 4 K_z K_{-}^2 + 2 K_{-}^2 \\
%    [K_{+}K_{z}K_{-},K_{+}^2] &= 2K_{+}^2K_{-} -4K_{+}^2K_{z}^2 \nonumber \\ &~~ - 6K_{+}^2K_z - 2K_{+}^2 . \\
%           [K_+^2 K_z,K_-^2] &= 8K_+K_z^2K_- - 2K_+^2K_-^2 \nonumber \\
%        &~+ 16  K_+K_z K_-  - 8K_z^3 \nonumber \\ &~+ 8K_+K_- - 4K_z^2 \\
%    [K_z^3,K_+^2] &= 6K_+^2K_z^2 + 12 K_+^2K_z + 8K_+^2
\end{align}
In terms of the operators normal ordered with respect to the mean field ground state, the commutator expressions are
   \begin{align}
       [\{K_z\},\{K_{\pm}^2\}] =& \pm 2 \{K_{\pm}^2\} \\
       [\{K_z^2\},\{K_{\pm}^2\}] =& \pm 4\{K_{\pm}^2K_z \} \pm 3\{K_{\pm}^2 \} \\
       [\{K_{+}K_{-}\},\{K_{\pm}^2\}] =&\mp 4\{K_{\pm}^2K_z\} \nonumber \\ &~ \pm 2(N\!-\!1)\{K_{\pm}^2\}  \\
       [\{K_{+}^2\},\{K_{-}^2\}] =& 8\{K_{+} K_{z}K_{-} \} \nonumber \\ &~+ (8\!-\!4N)\{K_{+}K_{-} \}
       - 8\{K_{z}^2\} \nonumber \\ &~+8(N\!-\!1)\{K_{z}\} \nonumber \\ &~- 2N(N\!-\!1)  \\
       %%%%%%%
       [\{K_{+}K_{z}K_{-}\},\{K_{+}^2\}] &= 2\{K_{+}^3K_{-}\} -4\{K_{+}^2K_{z}^2\} \nonumber \\
       &~+ 2(N\!-\!4 )\{K_{+}^2 K_{z}\} \nonumber \\
       &~+2(N\!-\!1)\{K_{+}^2\} . 
 %      %
%      [\{K_+^2 K_z\},\{K_-^2\}] &= 8\{K_+K_z^2K_-\} - 2\{K_+^2K_-^2\} \nonumber \\
%        &+ 4(5-N) \{ K_+K_z K_- \} - 8\{K_z^3\} \nonumber \\
%        &+ 8\{K_+K_-\} +8(N-2)\{K_z^2\} \nonumber \\
%        &- 2(N^2-5N+4)\{K_z\} \\
 %       %
%        [\{K_z^3\},\{K_+^2\}] &= 6\{K_+^2K_z^2\} + 9 \{K_+^2K_z\} + 3\{K_+^2\}
%       [\{K_{+}K_{z}K_{-}\},\{K_{+}^2\}] &= 2\{K_{+}^3K_{-}\} -4\{K_{+}^2K_{z}^2\} \nonumber \\
%       &~+ (4N-2)\{K_{+}^2 K_{z}\} \nonumber \\
%       &~~-N^2\{K_{+}^2\} - 6\{K_{+}^2K_{z}\} \nonumber \\
%       &~~~+3N\{K_{+}^2\} - 2\{[K_{+}^2\}. \nonumber \\
%%       &~~~+\tfrac{1}{2}N[\{K_{+}K_{-}\},\{K_{+}^2\}]
%       &~~~-2N\{K_+^2K_z\} +N(N-1)\{K_{+}^2\} 
   \end{align} 

%\srs{I suspect that some of the 3-body commutator terms are still wrong.}

\section{CCD solution to the LGM model}
The coupled cluster doubles solution of the LGM model is obtained from the decoupling equation
\begin{equation}
    \langle \phi | \{K_-^2\} e^{-T}H e^T |\phi \rangle = 0.
\end{equation}
where the CCD excitation operator is $T=t_2\{K_+^2\}$.
To solve this, we need the coefficient of $\{K_+^2\}$ in $H$, $[H,T]$, and $[[H,T],T]$.
Using the expressions in~\eqref{eq:LipkinCommNO}, we obtain
\begin{equation}
    \tfrac{1}{2} V + 2\epsilon t_2 + 2V(N^2-7N+9)t_2^2 =0
\end{equation}
defining $\alpha\equiv N^2-7N+9$, we obtain the solution
\begin{equation}
    t_2 = \frac{-\epsilon}{2V\alpha}\left[1-\sqrt{1-\alpha \left(\frac{V}{\epsilon}\right)^2 }  \right].
\end{equation}
We are interested in the region where $NV/\epsilon \sim 1$, so $t_2\sim \frac{N}{\alpha}\to \frac{1}{N}$ as $N\to \infty$.
The CCD correlation energy is
\begin{equation}
\begin{aligned}
    E_{CCD} & =\langle \phi | [H,T] \phi \rangle \\
    &= \tfrac{1}{2}Vt_2 \langle \phi |[\{K_-^2\},\{K_+^2\}]|\phi \rangle \\
    &= Vt_2 N(N-1)
    \end{aligned}
\end{equation}
which goes like $N$ as $N\to \infty$.

\section{Magnus(2) solution to the LGM model}

In the Magnus formulation of the IMSRG, the transformed Hamiltonian is given by~\eqref{eq:magnusH}.
At the NO2B level, the only off-diagonal operators that respect the symmetries of the problem are $K_{\pm}^2$.
The requirement of antihermiticity means $\eta(s)\propto (K_+^2-K_-^2)$ for all $s$, and so all the commutators in \eqref{eq:dOmegads} vanish.
Consequently, $\Omega$ is also proportional to this operator, and the transformation is specified by a single real number:
\begin{equation}
    \Omega = \tfrac{1}{2}\Omega_{\rm 2b}(K_+^2-K_-^2).
\end{equation}

Next we consider the behavior of nested commutators of $\Omega$ with $H$.
$[\Omega,H]$ yields one term proportional to $K_+^2+K_-^2$, and a second term proportional to 
\begin{equation}\label{eq:LGMOp1}
    [K_+^2-K_-^2,K_+^2+K_-^2]=2[K_+^2,K_-^2],
\end{equation}
so that
\begin{equation}\label{eq:LGMSingleComm}
    [\Omega,H] = a_1 [K_+^2,K_-^2]+b_1 (K_+^2+K_-^2).
\end{equation}
By applying the commutator expressions in~\eqref{eq:LipkinCommNO} we find in the NO2B approximation the double commutator
\begin{equation}\label{eq:LGMOp2}
    [K_\pm^2,[K_+^2,K_-^2]]=8 \tilde{\alpha} K_\pm^2
\end{equation}
with $\tilde{\alpha}\equiv N^2-5N+7$.
From \eqref{eq:LGMSingleComm}, \eqref{eq:LGMOp1} and \eqref{eq:LGMOp2}, we see that arbitrary nested commutators $[\Omega,\ldots[\Omega,H]\ldots]$ will have the same structure as \eqref{eq:LGMSingleComm}, so we may write for the $k$-fold nested commutator
\begin{equation}
    [\Omega,H]^{(k)} = a_k[K_+^2,K_-^2]+b_k (K_+^2+K_-^2).
\end{equation}
We have a recurrence relation for $a_k$ and $b_k$
\begin{equation}
    a_k = \theta^2 a_{k-2}~,\hspace{2em}
    b_k = \theta^2 b_{k-2}~
\end{equation}
with $\theta\equiv \sqrt{ 4\Omega_{\rm 2b}^2 \tilde{\alpha}}$,
and initial values
\begin{align}
    a_1 &= \tfrac{1}{2}\Omega_{\rm 2b} V \hspace{1em}  && a_2 = -\Omega_{\rm 2b}^2\epsilon  \\
 b_0 &=\tfrac{1}{2}V    &&b_1 = -\Omega_{\rm 2b}\epsilon .
\end{align}
Writing the off-diagonal piece of the transformed Hamiltonian as $\tilde{H}^{\rm od}=\tfrac{1}{2}\tilde{V}(K_+^2+K_-^2)$, we have
\begin{equation}
\begin{aligned}
    \tfrac{1}{2}\tilde{V} &= \sum_{k=0}^{\infty} b_k \frac{1}{k!}\\
%    &= \sum_{k~\rm{even}} \frac{\theta^k}{k!}b_0 + \sum_{k~{\rm odd}} \frac{\theta^k}{k!}b_1  \\
    &= \tfrac{1}{2}V\cosh\theta -\Omega_{\rm 2b}\epsilon \frac{\sinh\theta}{\theta} 
    \end{aligned}.
\end{equation}
The requirement that $\tilde{H}^{\rm od}\to 0$ gives the decoupling condition
\begin{equation} \label{eq:LGMDecopuling}
    \tanh \theta = \sqrt{\tilde{\alpha}}\frac{V}{\epsilon}
\end{equation}
which fixes $\Omega_{\rm 2b}$.
The ground state energy is obtained from the zero-body piece of $\tilde{H}$.
From the commutator expressions~\eqref{eq:LipkinCommNO} the zero-body piece of all nested commutator comes from $[K_+^2,K_-^2]_{\rm 0b}=-2N(N-1)$.
Consequently, the IMSRG(2) correlation energy is
\begin{equation}\label{eq:LGMEIMSRG}
\begin{aligned}
    E_{\rm IMSRG(2)} &= -2N(N-1)\left( \frac{V}{4\sqrt{\tilde{\alpha}}}\sinh\theta + \frac{\epsilon}{4\tilde{\alpha}}(1-\cosh \theta)\right)  \\
    &= -\frac{N(N-1)}{2\tilde{\alpha}} \left( 1 - \sqrt{1-\tilde{\alpha}\frac{V^2}{\epsilon^2}} \right).
\end{aligned}
\end{equation}
In the second line of \eqref{eq:LGMEIMSRG} we have used \eqref{eq:LGMDecopuling} and some identities of hyperbolic functions.
%The amplitude equation can also be solved by iteration
%\begin{equation}
%\frac{V}{4\epsilon} + t_2^{[n+1]} + \frac{V\alpha}{\epsilon} t_2^{[n+1]} t_2^{[n]}=0
%\end{equation}
%or
%\begin{equation}
%    t_2^{[n+1]} = -\frac{V}{4\epsilon} / (1 + \frac{V\alpha}{\epsilon} t_2^{[n]} )
%\end{equation}
%\begin{equation}
%    t_2^{[n+1]}-t_2^{[n]} = \frac{-V/4\epsilon - (1+\alpha V/\epsilon t_2^{[n]})t_2^{[n]}}{1 + \alpha V/\epsilon t_2^{[n]}}
%\end{equation}

\bibliography{references}% Produces the bibliography via BibTeX.

\end{document}